\documentclass[reprint,amssymb,amsmath,aip,cha]{revtex4-2}
\usepackage{graphicx}
\usepackage[caption = false]{subfig}
\usepackage{color}
\usepackage{epsfig}
\usepackage{ifpdf}
\usepackage{bm}
\usepackage[colorlinks=true,linkcolor=blue]{hyperref}%
\expandafter\ifx\csname package@font\endcsname\relax\else
 \expandafter\expandafter
 \expandafter\usepackage
 \expandafter\expandafter
 \expandafter{\csname package@font\endcsname}%
\fi
\usepackage{cleveref}
\begin{document}
\title{Degradation of methylene blue dye as wastewater pollutant by atmospheric pressure plasma: A comparative study}
\author{Chintan Vyas} 
\affiliation{Institute of Advanced Research, Koba, Gandhinagar, 382426, Gujarat, India}
\author{Mangilal Choudhary}
\email{mangilalchoudhary@banasthali.in}
\affiliation{Banasthali Vidyapith, Tonk, 304022, Rajasthan,India}
%
\begin{abstract}
The industrial application of non-thermal plasma has been a research field in the last few years. One of the potential applications of non-thermal plasma is in treating dye effluents of textiles industries which are considered as one of environmental pollutants. Before scaling up plasma technology at the industrial level, it is required to understand the interaction of non-thermal plasma with a synthetic dye-containing solution in laboratory experiments. A detailed comparative study of MB dye degradation using an atmospheric pressure air plasma (corona discharge) source is carried out in this report. Different concentrations (5 mg/L, 10 mg/L, and 40 mg/L) of MB solutions were treated with plasma and were observed various chemical parameters (pH, TDS, EC, etc.) as well as degradation percentages. We observed nearly 80 \% MB degradation in about 100 min, 110 min, and 130 min treatment times for 5 mg/L, 10 mg/L, and 40 mg/L MB solutions respectively. With increasing the volume of the MB sample solution (15 ml to 30 ml), nearly double the plasma treatment time is required to decompose the same amount of MB dye. We observed a higher degradation percentage ($\sim$ 90 \%) of MB solution (10 mg/L) in 100 minutes when iron material (sheet) is used as cathode. However long time treatment (170 to 180 minutes) to degrade $\sim$ 90 \% MB dye in the case of copper or aluminium cathode is reported. The MB degradation percentage can be increased from $\sim$ 60 \% to 90 \% if the pH of the MB solution is changed from mild basic (pH $\sim$ 8.5) to acidic (pH $\sim$ 3.5) in the case of the aluminium cathode in nearly 60 min. The pH of the MB solution is changed by adding low pH treated dye solution instead of any external chemicals. The results show that acidic MB solution takes less time, about 90 minutes instead of 180 minutes to degrade approximately 90 \% MB dye. The pH of the solution decreases during plasma treatment but the TDS and EC of MB solution increases during plasma treatment irrespective of dye concentration, the volume of solution, and different types of cathodes (iron, copper, and aluminium). The results are qualitatively discussed in line with the available theoretical and experimental background of plasma-water interaction.
\end{abstract} 
\maketitle
\section{Introduction}
Large-scale manufacturing and applications of synthetic reactive dyes in textile, rubber, plastic, cosmetic and pharmaceutical industries have been a source of worry since it is a major source of water pollution. A fraction of dye which is not used in the process ends up in wastewater generated by dye-using industries. It is fact that only a small portion of the generated wastewater is adequately treated using various techniques. The rest of industrial wastewater is usually dumped or directly discharged into the sewage system, which ultimately enters rivers, ponds, lakes, etc.
Such low concentrations of organic compounds or dyes in the discharged wastewater into natural streams (e.g. rivers, ponds, etc.) have been a cause of many significant problems such as increasing toxicity, reducing light penetration, affecting the health of living things (plants, animals, and humans), the origin of various diseases, etc. It has been a priority for environmental researchers to remove dyes and other organic compounds (toxic) from industrial wastewater to reduce adverse effects on the environment. A wide spectrum of research has been conducted to minimize the environmental pollution caused by effluents released by the textile and other chemical or dye industries. There are various techniques such as physical, biological, and chemical used to degrade the organic dyes or remove organic contaminants present in the dye-using industrial wastewater \cite{physicalchemicalmethode2, chemicalmethod2, biologicalmethod1,biologicalmethod2,biologicalmethod3}. \\
Physical methods such as flotation, filtration, adsorption, etc. are non-destructive physical separation processes. These processes are generally efficient to remove the dyes or organic compounds from wastewater but they only transfer pollutants (organic compounds or dyes) from the liquid phase to the solid phase. Therefore, post-treatment of toxic solid waste and regeneration of the adsorbent materials are required in these techniques \cite{physicalchemicalmethode2} Biological treatment is also a promising technique to degrade the dye molecules but the treatment processes are very slow and need experts to handle the reactors. The high toxic level or pH of the wastewater also restricts the biological treatment (biological reactions) for the dyes decomposition \cite{biologicalmethod1,biologicalmethod2,biologicalmethod3,physicalchemicalmethode2}. Chemical methods are also widely used to degrade the dye molecules but there is a requirement of various chemical agents as well as safety measures to handle the chemicals \cite{chemicalmethod2,physicalchemicalmethode2} during treatment. We know that organic compounds or dyes have very complicated aromatic structures, therefore, biological and chemical oxidation processes (techniques) are ineffective for the degradation of these chemical compounds in wastewater. The studies on dye or organic complex compound degradation using various methods confirmed that the oxidation process is considered one of the most effective emerging techniques for the treatment of wastewater containing chemical pollutants (dyes or other organic compounds). In the oxidation process, either oxidizing agents ($H_2 O_2$, $O_3$)  or reactive oxidant species (hydroxyl radical ($\dot{OH}$) under suitable conditions interact with organic pollutants and decompose them in the wastewater. In the last few years, the potential application of advanced oxidation processes (AOPs) in the degradation of harmful organic complex compounds in industrial wastewater has been a hot topic of research. In the advanced oxidation process, various reactive oxygen species along with UV radiation and catalysts help in the decomposition of organic compounds which is difficult in simple oxidation approaches \cite{advancedoxidation1,advancedoxidation2,advancedoxidation3,advancedoxidation4,advancedoxidation5,advancedoxidation6}.

The advanced oxidation processes are more costly due to the use of expensive reagents (e.g., $H_2O_2$) and energy consumption (In the generation of $O_3$ or UV radiation). Therefore, some new Eco-friendly low-cost advanced oxidation processes are required in decomposing organic compounds. \\ 
In recent years, low-temperature plasma technology which is considered an advanced oxidation technique has been investigated for the degradation of organic compounds in water or wastewater \cite{wasterwaterplasma1,wastewaterplasma2,wastewaterplasma4}. Compared to other discussed advanced oxidation processes, the low-temperature plasma-based oxidation technique can simultaneously generate various oxidative agents such as UV radiation, energetic electrons, $H_2O_2$, $O_3$, and other reactive species \cite{plasmawaterinteractionreview1,plasmawaterinteractionreview2,rosspecieswater1,wasterwaterplasma1,wastewaterplasma4}. It is possible to generate all these oxidant agents (oxygen reactive species) without using any chemical agent. Therefore, it can be considered a green, eco-friendly, and cost-effective technique to decompose wastewater containing organic compounds or dying wastewater. The non-thermal (low-temperature) air or nitrogen-oxygen ($N_2/O_2$) plasma contains various reactive oxygen and nitrogen species such as hydrogen peroxide ($H_2 O_2$), hydroxyl radical ($\dot{OH}$), singlet oxygen ($1O_2$), oxygen radical ($\dot{O}$), nitric oxide ($NO$) and nitrite/nitrate anions ($NO_2^{-} / NO_3^{-}$), various ions, etc. In the gaseous phase, non-thermal plasma (NTP) is a source of visible radiations, UV radiations, energetic electrons, and long-lived stable molecules ($O_3$) along with reactive nitrogen and oxygen species \cite{rosspecieswater1,wastewaterplasma4,plasmawaterinteractionreview1}. If non-thermal (corona discharge) air plasma (at atmospheric pressure) interacts with water or wastewater then plasma reactive species, $O_3$, UV radiation, and energetic electrons interact with water (wastewater) and strong oxygen-based oxidizers, such as hydroxyl radicals ($\dot{OH}$), ozone ($O_3$), hydrogen peroxide ($H_2O_2$), and oxygen radical ($\dot{O}$) are expected to form in the water solution. These advanced active oxidizing agents significantly contribute to the degradation of organic contaminants or reactive dyes in industrial wastewater \cite{wasterwaterplasma1,wastewaterplasma2,wastewaterplasma4,mbdegradation1,mbdegradation2,mbdegradation3,mbdegradation4}. The interaction of UV radiations or energetic electrons can also play a significant role in decomposing the organic complex compounds along with the oxygen-based oxidisers. Thus the plasma-based advanced oxidation process (AOP) has great potential to decompose the organic contaminants in wastewater. In combination with a physical or biological wastewater treatment approach, Non-thermal plasma (NTP) has a great potential to treat wastewater containing various organic reactive dyes or other organic compounds at low effective cost. It is also expected to form various reactive nitrogen species such as nitric oxide ($NO$), nitrite anions (acid) ($NO_2^{-}$), nitrate anions ($NO_3^{-}$), etc. during the water-plasma interaction. The chemical parameters of plasma-treated wastewater or water like pH, electrical conductivity, etc. are also controlled by dissolved nitrogen reactive species. 
 \\
For scaling up the plasma technology at the industrial level, a detailed study of organic reactive dye degradation in the solution while plasma interacts at the water surface is required. In earlier reported works, researchers used hydrochloric acid (HCL), sulphuric acid ($H_2SO_4$), or sodium hydroxide (NaOH) to control the pH of MB dye solution to explore the effect of pH on the efficiency of MB degradation \cite{pheffectonmb}. The past study suggests that Fenton's reaction increases the efficiency of MB degradation. They used ferric chloride ($FeCl_3$) or other $F_e$ compounds to initiate the Fenton reaction \cite{fantonreaction1,fantonmbdegradation1, fantonmbdegradation2}. After going through past research works on dye degradation using low-temperature plasma sources, we ended up with some open questions about MB dye degradation. Is it possible to reduce the pH of dying wastewater (MB solution) without adding any external chemicals? Could it be possible to increase the efficiency of MB dye degradation without using any ferrous compounds (chemicals)? Is there a linear relation between dye degradation rate and dye concentration in the case of plasma treatment? To get answers to these questions, we performed experiments using air plasma (spark glow discharge) with a model system (MB dye solution) which is assumed to be a wastewater sample containing synthetic dye. \\
In the present report, different sets of experiments were performed to explore the influence of different factors such as MB dye concentration, the volume of solution, pH of the solution, cathode materials, etc. on the efficiency of MB degradation with atmospheric air plasma (corona discharge) treatment. We performed a few experiments with the iron cathode  without using any chemicals to predict the effect of cathode material (iron) on the efficiency of MB dye degradation. We only used plasma-treated MB dye solution (colorless) instead of any chemical (acid) to make the MB solution more acidic before plasma treatment. The experimental results are explained based on the available theoretical model and experimental results of plasma-water interaction. A detailed description of the experimental set-up and techniques is given in Sec.\ref{sec:exp_setup}. The experimentally observed results are discussed in Sec.\ref{sec:results_discussion}. A brief summary of the work along with concluding remarks is provided in Sec.\ref{sec:summary}.
\section{Experimental Setup and Methods} \label{sec:exp_setup}
In the present study, we used a commercially available high voltage ($V_{p-p}$ = 6 kV) and low current ($<$ 1 A) to treat the model solutions of methylene blue dye (MB-dye). There was a provision on the power supply to set the on and off time of sparking between the high voltage (H.V.) electrode (anode) and the grounded electrode (cathode). The high melting point alloy wire of diameter 3 mm as H.V. anode and rectangular shaped (20 mm $\times$ 15 mm) grounded cathode of different materials (Aluminium, Iron, and Copper) were used to ignite the spark glow (corona discharge) at atmospheric pressure. To demonstrate the role of plasma species in the decomposition of complex dyes or organic molecules, the study was conducted with the Methylene blue dye (MB dye) for the various parameters. The MB dye solutions of different concentrations were prepared using the appropriate volume of simple tap water. A Thermometer was used to measure the temperature of plasma treated MB solution at different treatment times. The electrical conductivity (EC) of MB dye solution measures its capacity to conduct the electrical current due to dissolved salts and other inorganic/organic compounds. An electrical conductivity meter was used to measure the EC of solutions in siemens per meter ($S/m$). The pH of aqueous or MB dye solutions is a quantitative measure of their acidity or basicity. The MB dye solutions after plasma treatment were classified as acidic or basic based on the pH value. A pH meter was used to measure the pH of untreated and plasma-treated MB dye solutions. The amount of total dissolved solids (TDS) in untreated and plasma-treated dye solutions was measured using the TDS meter in parts per trillion (ppt). Apart from these parameters of dyes, a UV-Vis spectrometer in of range 200 nm to 800 nm was used to get the percentage of MB degradation with plasma treatment.   
\begin{figure*} 
\subfloat{{\includegraphics[scale=0.40050]{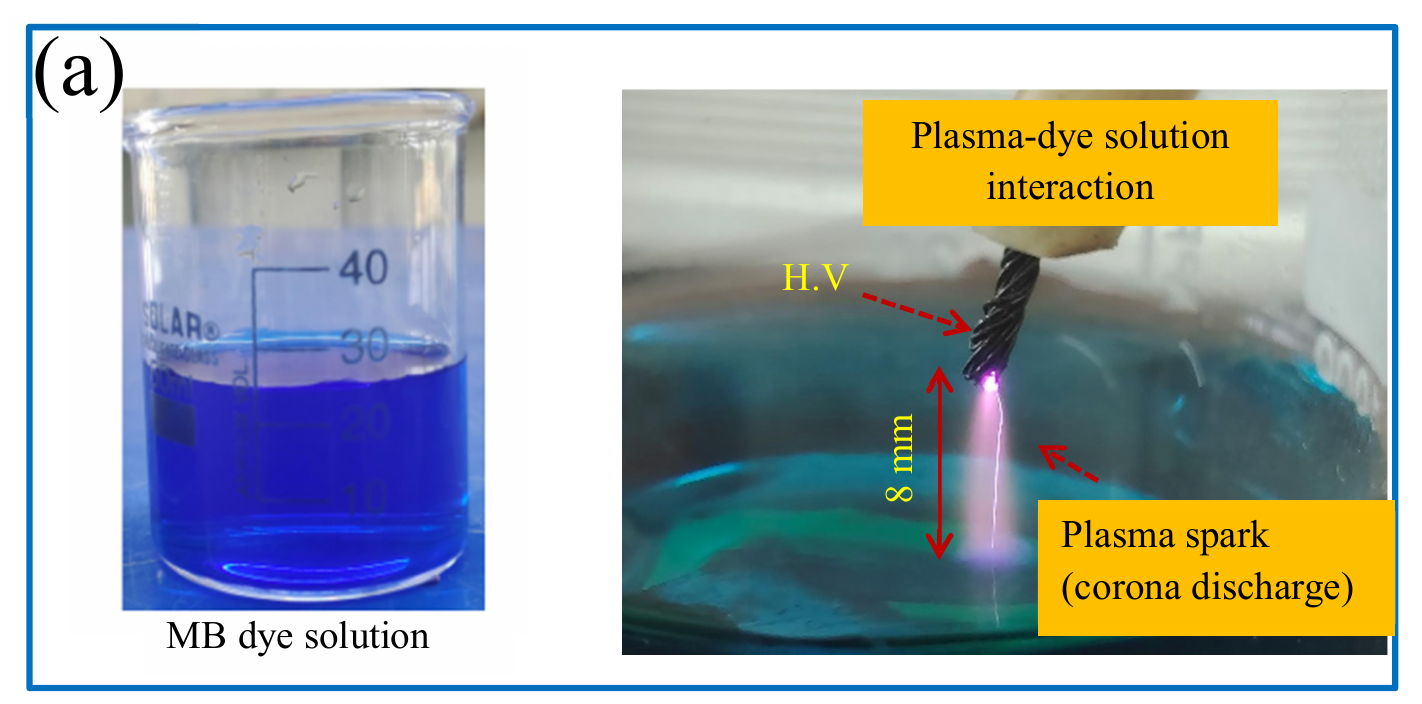}}}%
\hspace*{0.2in}
 \subfloat{{\includegraphics[scale=0.40050]{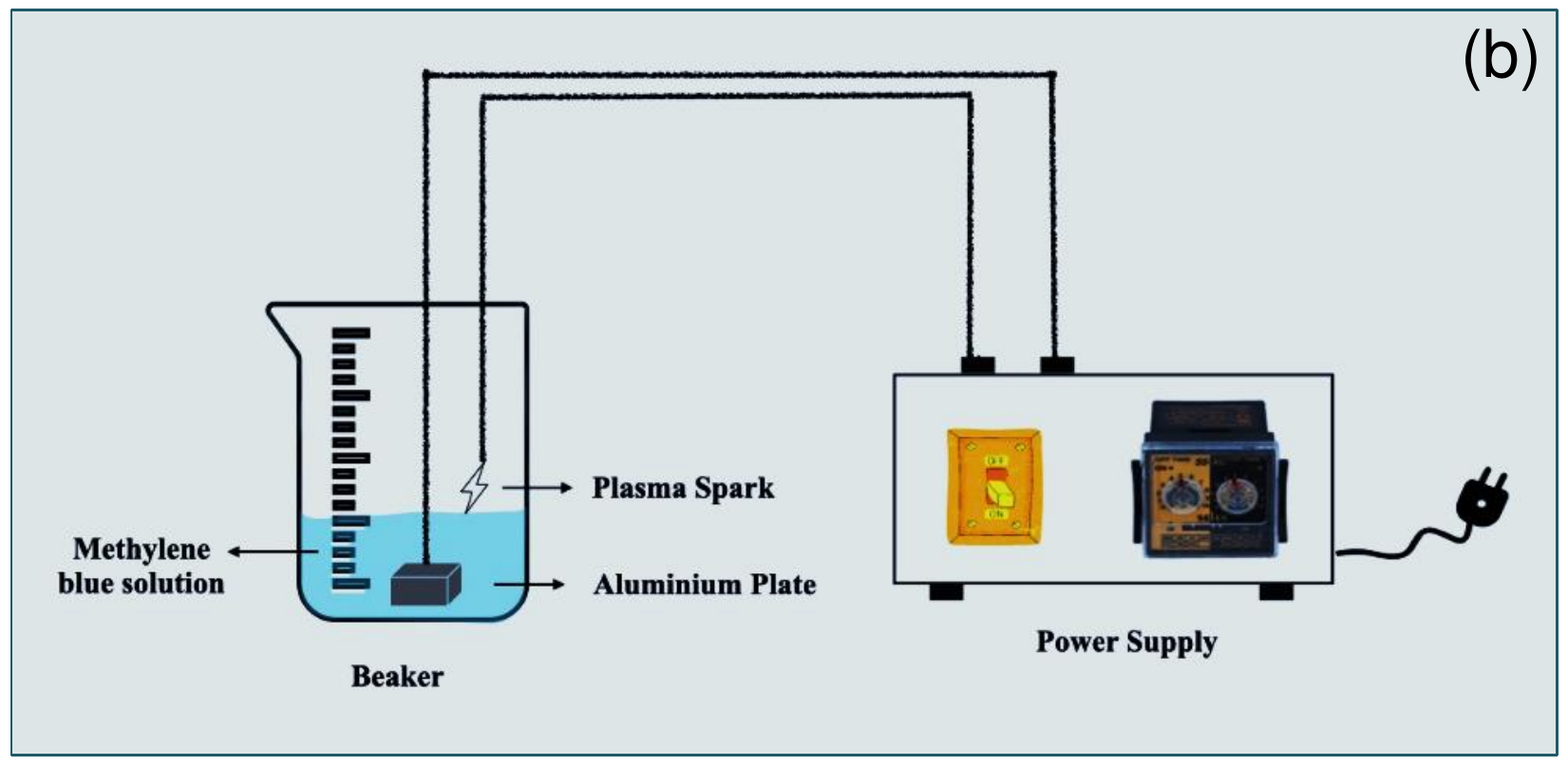}}}
\caption{\label{fig:fig1}(a) MB dye solution treatment by atmospheric pressure air plasma (b) Schematic diagram of the experimental setup used in treating the MB dye solutions} 
\end{figure*}
\begin{figure*} 
\subfloat{{\includegraphics[scale=0.320050]{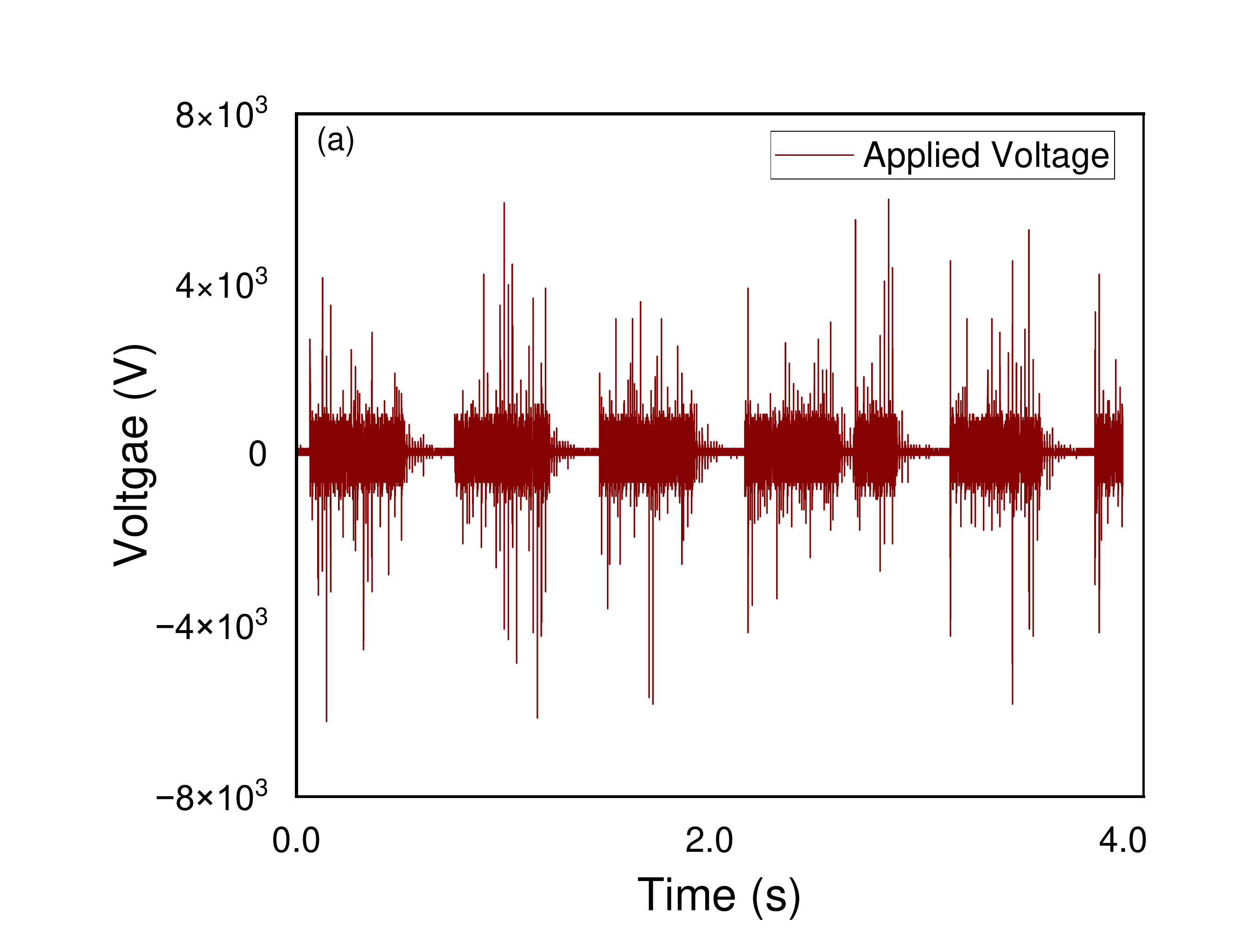}}}%
 \subfloat{{\includegraphics[scale=0.320050]{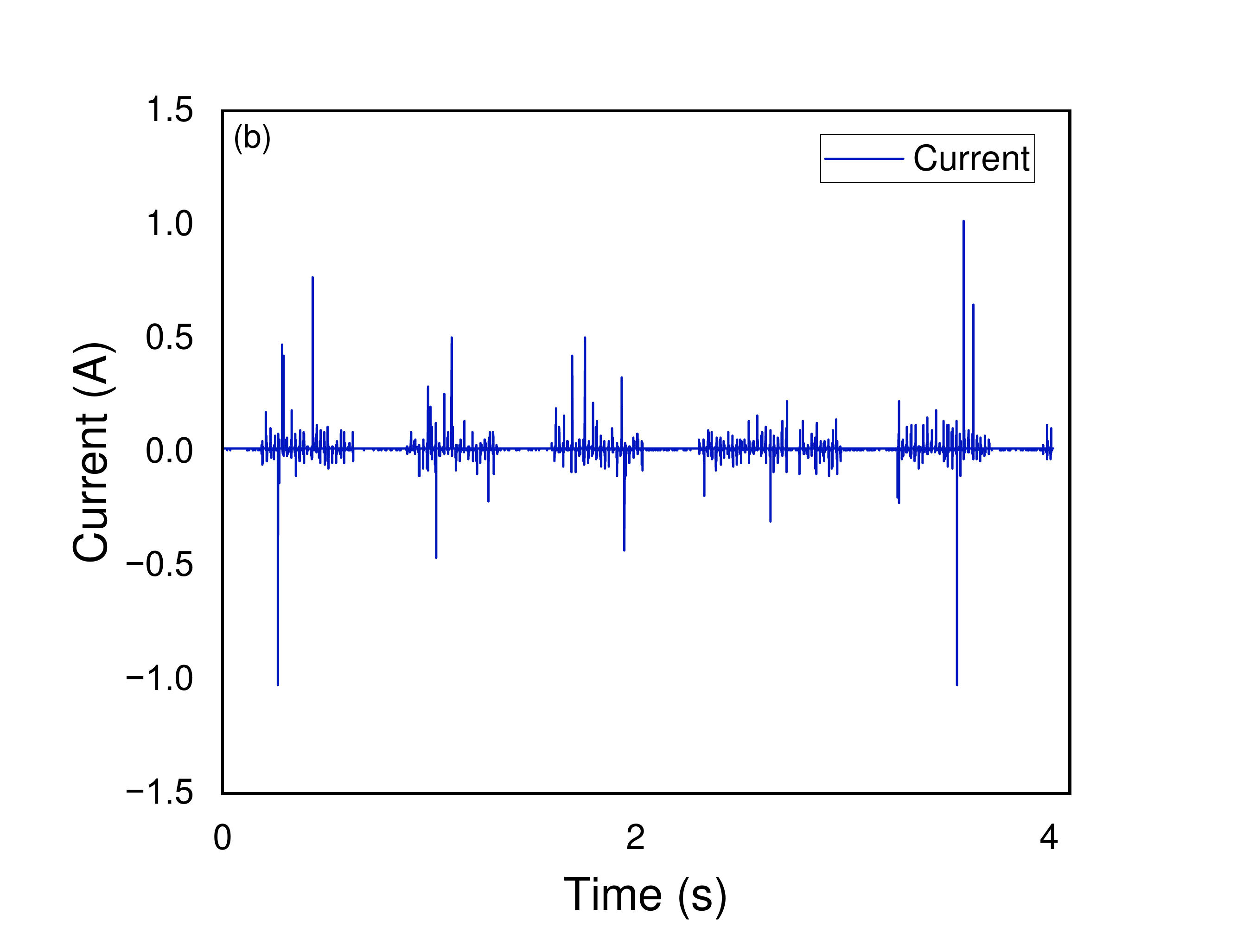}}}
\caption{\label{fig:fig2}(a) Voltage pulses (b) Current pulses} 
\end{figure*}
\section{Experimental Observations} \label{sec:results_discussion}
For a detailed comparative study of MB dye degradation with the atmospheric air plasma (corona discharge) treatment, a set of experiments was performed for different dye concentrations, various volumes of MB dye solutions, different electrode materials, various pH values of MB solutions, etc. The experimental conditions and outcomes of various experiments are discussed in subsequent subsections.
\subsection{Effect of MB dye concentration on MB degradation}
\begin{figure} 
\centering
 \includegraphics[scale= 0.4500000]{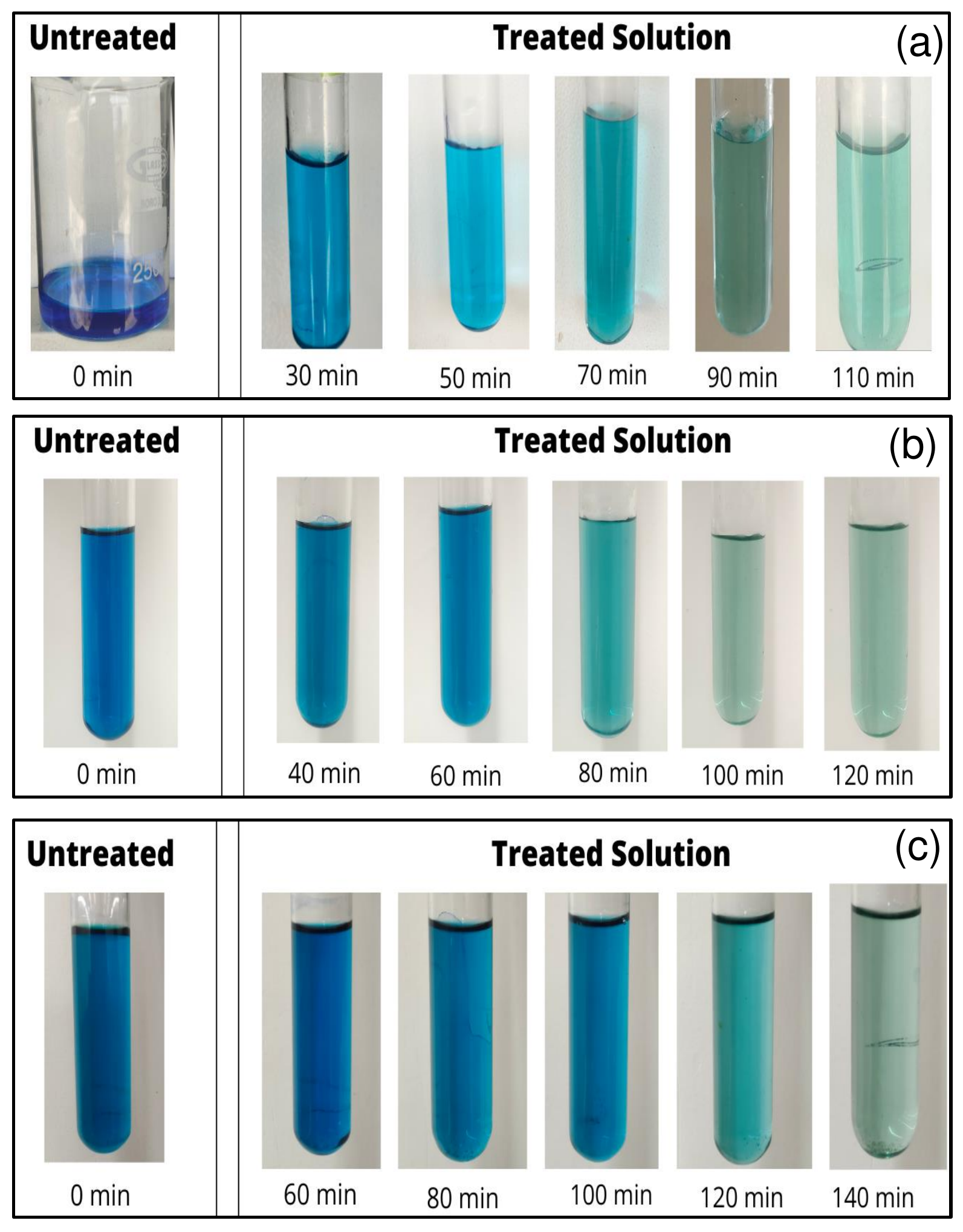}
\caption{\label{fig:fig3} Images of MB dye solutions (untreated and plasma treated) of different concentrations (a)5 $mg/L$ (b)10 $mg/L$ and (c)40 $mg/L$ MB dye solutions at different treatment times}
\end{figure}

\begin{figure*} 
\subfloat{{\includegraphics[scale=0.30050]{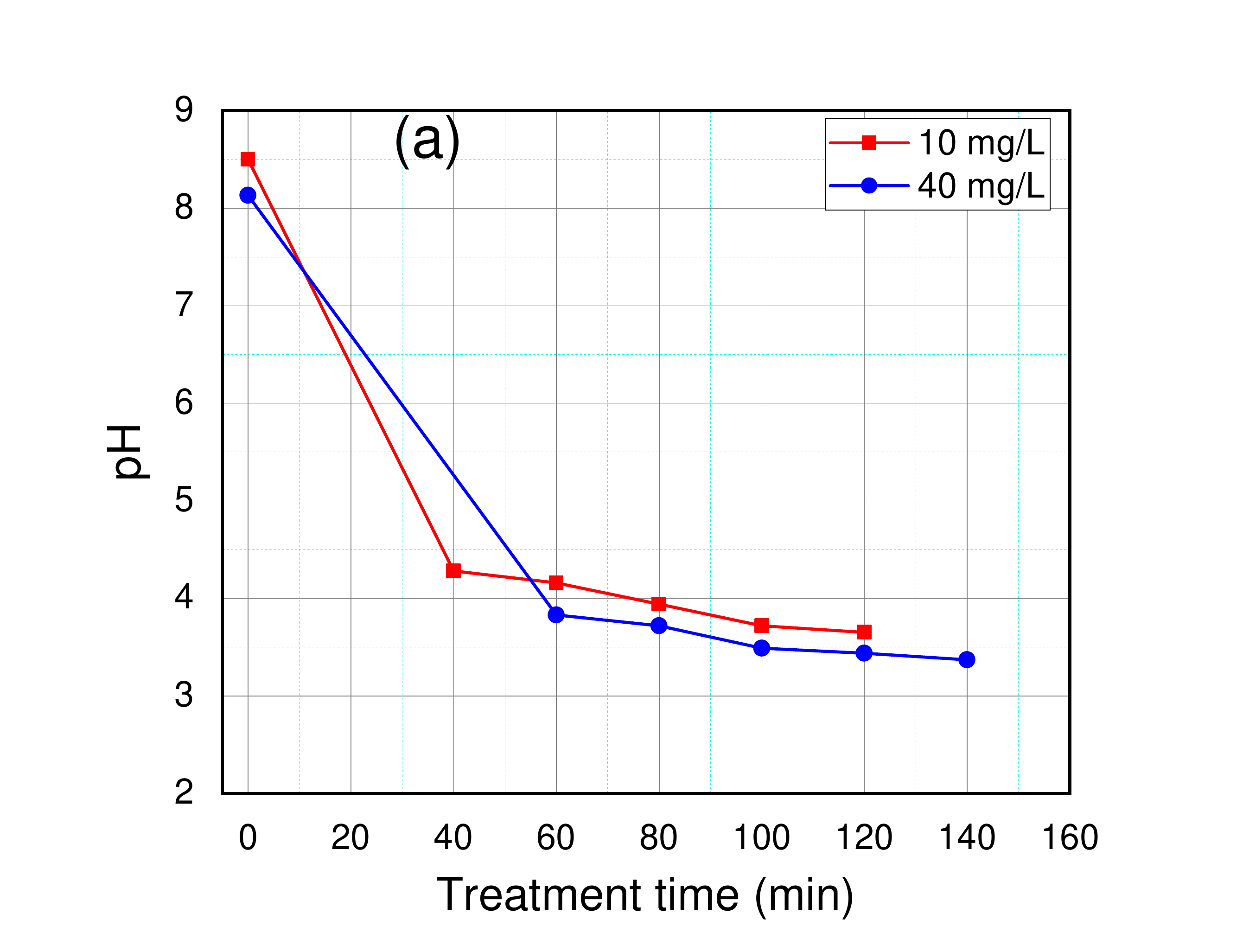}}}%
 \subfloat{{\includegraphics[scale=0.30050]{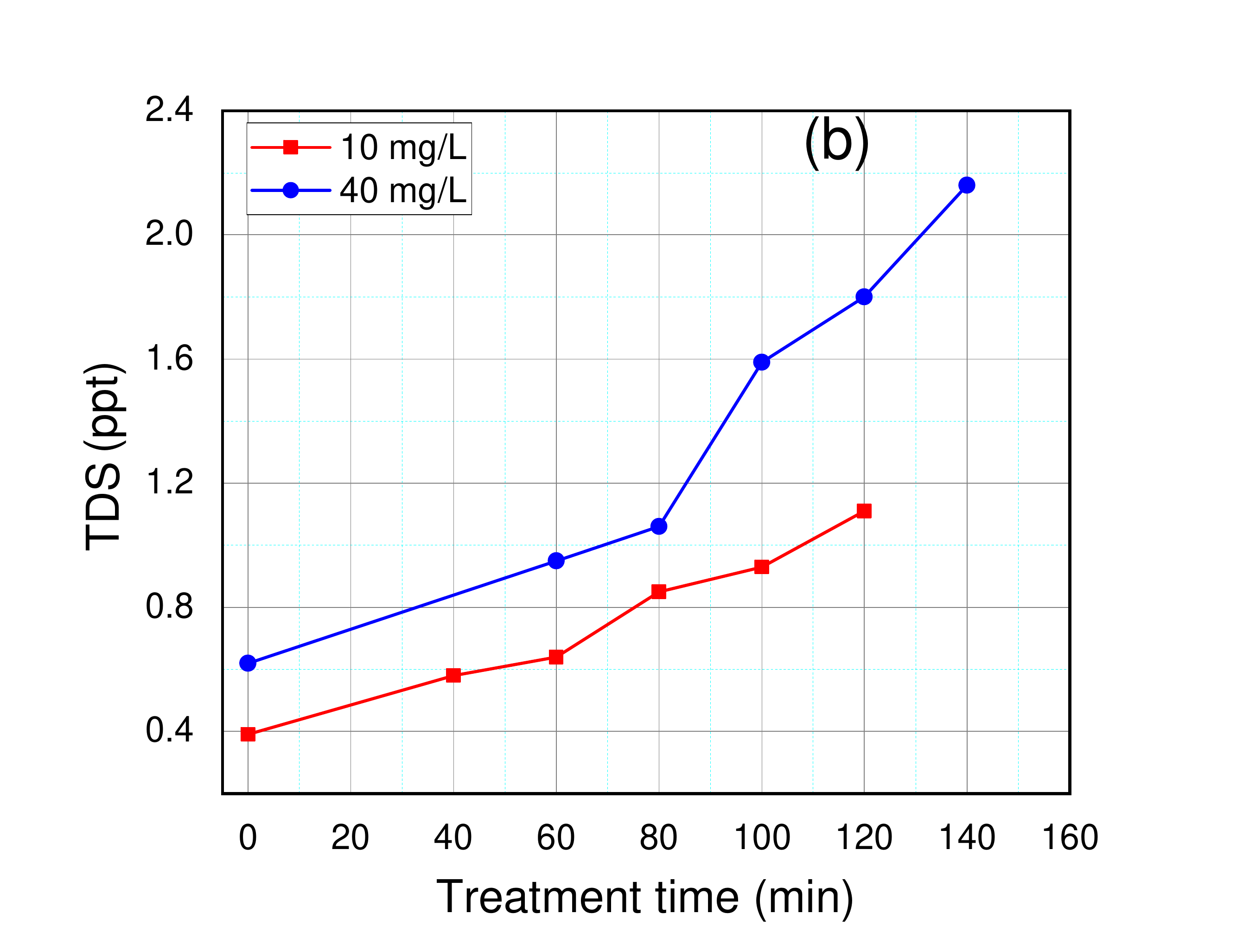}}}
 \qquad
 \subfloat{{\includegraphics[scale=0.30050]{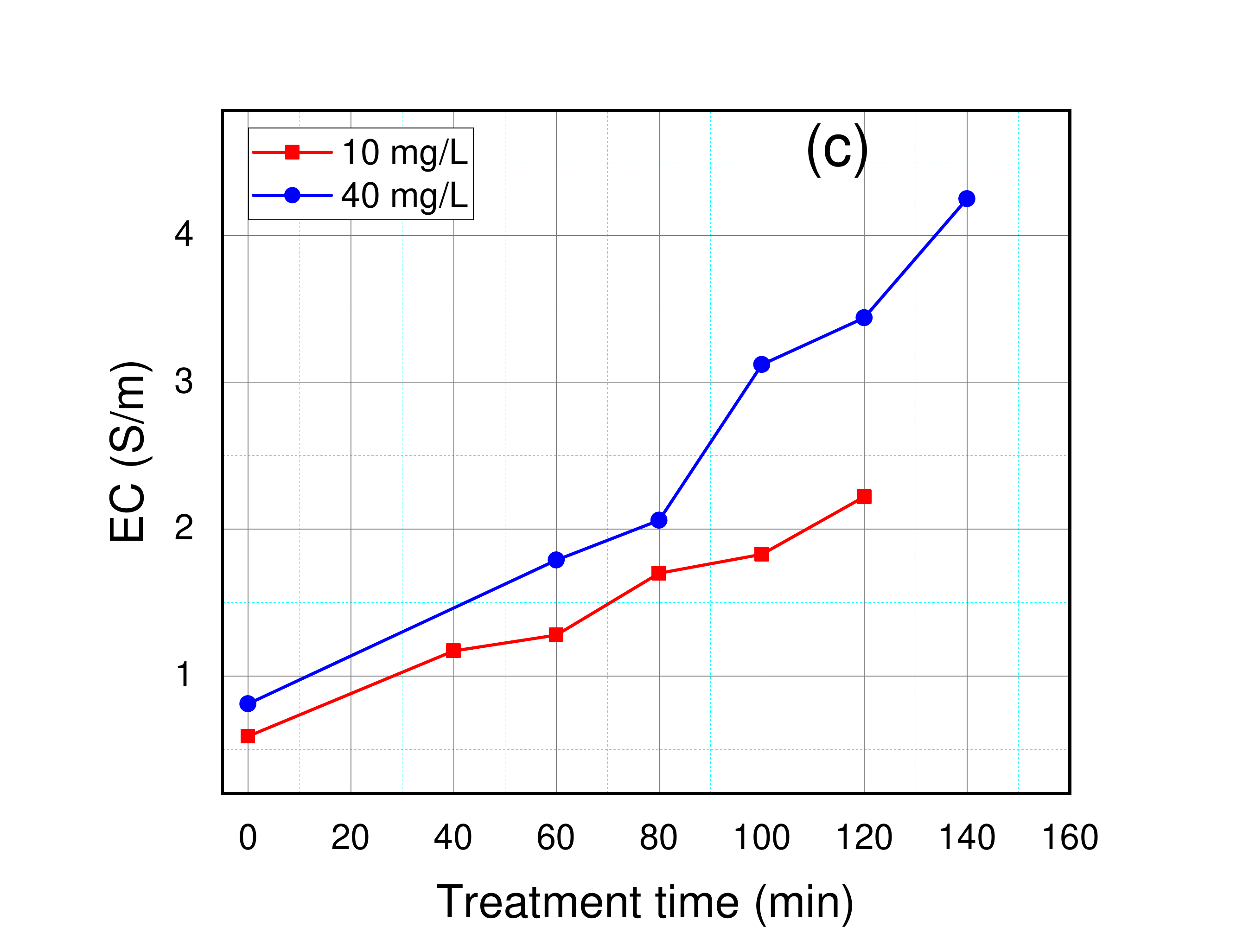}}}
\caption{\label{fig:fig4}(a) pH, (b) Tds, and (c) EC variation for two different concentrations (10 $mg/L$ and 40 $mg/L$) MB dye solutions. Error over the averaged value of the measured parameters for untreated and plasma-treated dye solutions are $< \pm$ 5\%.} 
\end{figure*}

In this set of experiments, MB dye solutions of different concentrations (5 $mg/L$, 10 $mg/L$, and 40 $mg/L$) were treated with the non-thermal plasma (NTP). 15 $ml$ MB solution was used as a sample for the plasma treatment. The same volume of samples was treated at different times and stored in a dark place for further chemical composition analysis. To avoid the effect of photons/lights on dye degradation, experiments were conducted in a lab after making it a kind of dark room. The change in color of MB dye solutions during plasma treatment can be seen in Fig.~\ref{fig:fig3}. The MB color changes from dark blue through light blue to nearly colourless in approximately 150 min. The change in color of the MB solution after treatment with non-thermal plasma (corona discharge) is due to the reduction of the number of MB dye molecules in the solution. It indirectly indicates the degradation of MB molecules during the plasma interaction.\\
In Fig.~\ref{fig:fig4}, we have plotted some parameters of MB solutions after treatment with the air plasma. The pH of MB dye solution against plasma treatment time plot (Fig.~\ref{fig:fig4}(a)) clearly
shows a fast drop in pH value during initial plasma treatment (up to 40 to 50 min) and then decreases with a slow rate with increasing the plasma treatment time. The lower value of pH is an indication of the acidic behavior of the MB solution. However, the 40 mg/L sample is more acidic than the 10 mg/L samples at the same treatment time. The variation of  total dissolved solids (TDS) and electrical conductivity (EC) is almost linear with plasma treatment time. Both TDS and EC are increasing with plasma treatment time but the amount of TDS in 10 mg/L is found less than 40 mg/L MB solution as shown in Fig.~\ref{fig:fig4}(b) and Fig.~\ref{fig:fig4}(c). Higher TDS and EC are a signature of a higher rate of MB degradation at a given time in different concentrations of MB dye solutions.
\begin{figure} 
\centering
 \includegraphics[scale= 0.3300000]{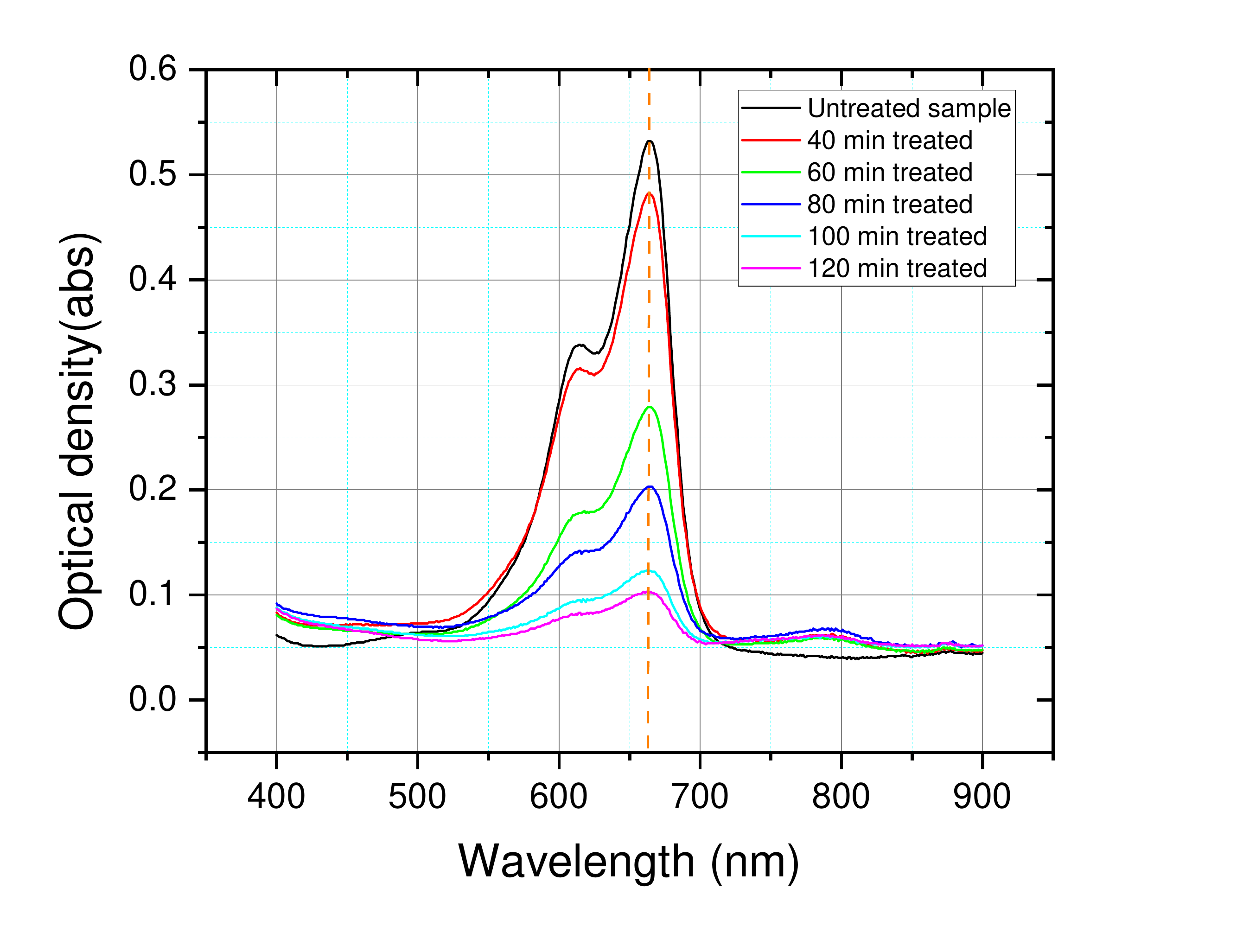}
\caption{\label{fig:fig5} UV-Vis absorption spectra for 10 $mg/L$ concentration MB dye solution.}
\end{figure}
It is also required to know the MB degradation percentage (efficiency of MB degradation) with air plasma treatment, therefore, samples were further analyzed using a UV-VIS absorption spectrophotometer (100 to 800 nm wavelength range). UV-VIS spectrophotometry is a technique for measuring light absorption in the ultraviolet and visible range of the electromagnetic spectrum. The MB dye molecule has a tendency to absorb UV and Visible bands of the spectrum. Based on the intensity of absorbed light of a particular wavelength by MB dye molecules (organic molecules), it is possible to estimate the percentage (\%) of MB dye molecules present in the solution at that time. 
In the present study, MB dye quantity during the plasma treatment is characterized by an absorption peak at 664 nm. The conversion (degradation) is calculated as conversion (\%) = 100 $\times$ actual concentration (height of absorption peak)/initial concentration (maximum height of absorption peak). The absorption spectra of MB dye solution (10 mg/L) at different plasma treatment times are depicted in Fig.~\ref{fig:fig5}. It should be noted that absorption spectra of different concentrations of MB dye solution (5 mg/L and 40 mg/L) were also analyzed for a comparative study of MB degradation during the plasma treatment. \\
\begin{figure} 
\centering
 \includegraphics[scale= 0.3200000]{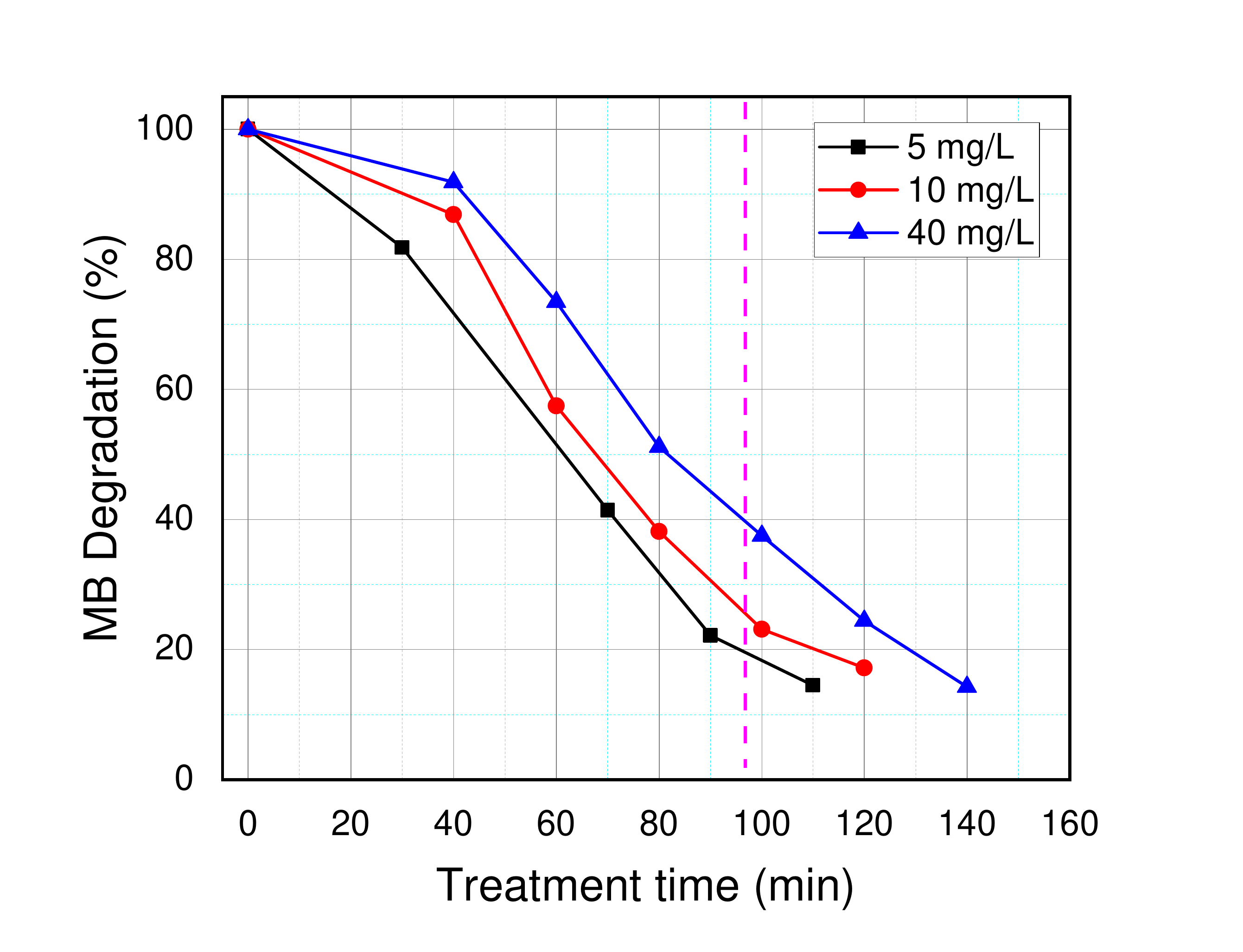}
\caption{\label{fig:fig6} Degradation (in \%) of different concentrations (a)5 $mg/L$ (b)10 $mg/L$ and (c)40 $mg/L$ MB dye solutions with plasma treatment. Error over the plotted values (data) is $< \pm$ 5\%.}
\end{figure}
The variation of MB degradation in percentages of 5 mg/L, 10 mg/L, and 40 mg/L solutions at different plasma treatment times is shown in Fig.~\ref{fig:fig6}. Before the treatment (at t = 0), the concentration of MB molecules is 100 \% in the sample solution. As plasma treatment start, the concentration of dissolved MB dye molecules in the solution is reduced (see Fig.~\ref{fig:fig6}). At a given treatment time (t = 98 min), we observed approximately 60 \%, 75 \% and 82 \% MB degradation in 40 mg/L, 10 mg/L and 5 mg/L solution respectively. It shows that a longer plasma treatment is required to degrade MB dye if the density of MB dye molecules in a given volume is increased. It is also observed that the rate of MB degradation is higher ($>$ 70 \%) at initial treatment (up to 80 to 90 min) and then degradation takes a longer time. There is a small difference ($<$ 15 \%) in degradation percentage for 10 mg/L and 40 mg/L (see Fig.~\ref{fig:fig6}) even though the density difference of MB dye molecules is four times. It can be concluded that degradation time does not have a linear relation with MB dye density in a given treated sample. A longer plasma treatment is required to degrade dye molecules (organic compounds) in a dense (high-concentration) solution than a dilute (low-concentration) solution.  
\subsection{Effect of Volume on MB degradation}
\begin{figure} 
\centering
 \includegraphics[scale= 0.450000000]{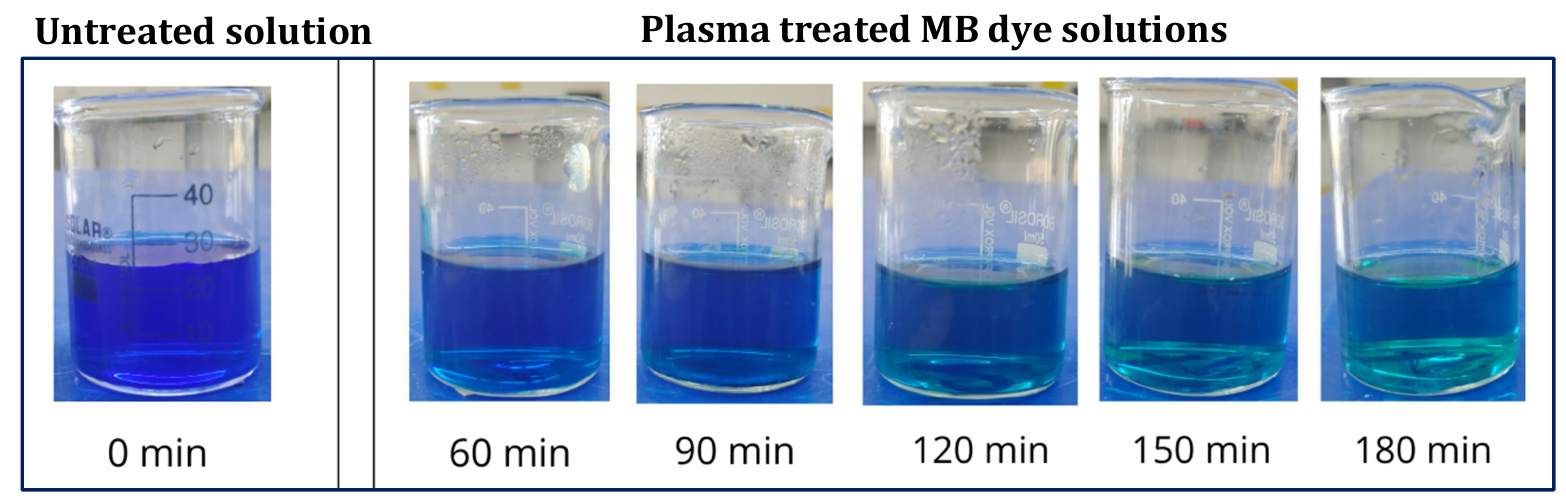}
\caption{\label{fig:fig7} Images of MB dye solution of volume 30 ml and concentration 10 $mg/L$ at different plasma treatment times}
\end{figure}

\begin{figure} 
\centering
 \includegraphics[scale= 0.3200000]{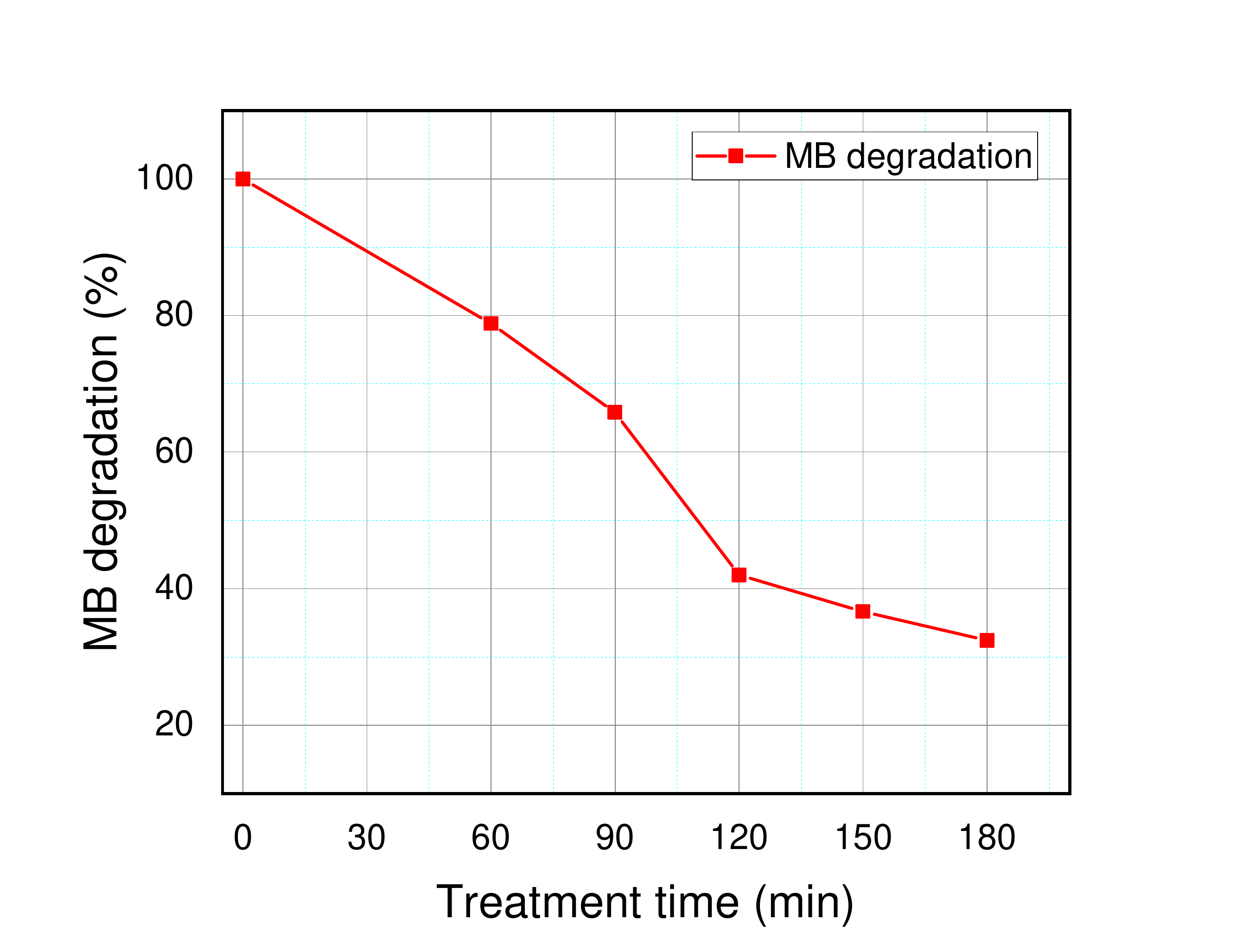}
\caption{\label{fig:fig8} Degradation (in \%) of 10 $mg/L$ concentration MB dye solution (30 ml) with plasma treatment time. Error over the plotted values (data) is $< \pm$ 5\%.}
\end{figure}

After studying the MB degradation in different concentrations of the solution, our focus was to see the effect of the volume of MB dye solution on degradation time with the same plasma source. A set of experiments was performed with MB solution (10 $mg/L$) of double volume (30 ml) compared to the previous study (15 ml). The 30 ml samples were treated at different times and stored in a dark place for further chemical composition analysis.
\begin{table*}
\caption{Chemical parameters of plasma treated MB dye solutions (volume = 30 ml) at a different time when the cathode was an aluminum plate.}
\centering 
\begin{tabular}{|l|c|c|c|c|c|c|c|c|c|c}
\hline
\hline 
Treatment time& Temperature & pH & EC & TDS \\
 (min) &($C^0$)&  & (S/m) & (ppt)\\
\hline
0& 30 & 8.6 & 0.66 & 0.33\\   
 60 & 49& 4.31 & 1.03& 0.51 \\
  90& 56&  4.16 & 1.15& 0.58 \\   
  120& 57& 3.8 &1.51 & 0.76 \\
  150 & 57& 3.3 &1.73 & 0.86 \\
  180 & 58& 2.56 & 2.45 & 1.24 \\
\hline \hline
\end{tabular}
\end{table*}
 %
The color images of MB dye solutions which were treated by plasma at different times are displayed in Fig.~\ref{fig:fig7}. We notice a change in color if plasma treatment time is increased. It predicts the reduction of MB dye molecules in a solution after plasma treatment. The measured parameters such as temperature, pH, TDS, and EC of plasma-treated MB samples are given in Table 1. The solution temperature rises first and then remains nearly constant during the plasma treatment. The pH of the plasma-treated solution decreases and the solution becomes acidic during plasma treatment. We also observe the increase in TDS and EC values of MB solution after treating it with plasma. For a further comparative study, we plotted MB degeneration percentages with plasma treatment time in Fig.`\ref{fig:fig8}. At t = 0, the solution has 100 \% MB molecules but after plasma treatment, we see a reduction in MB dye molecules. The degradation rate of MB molecules is higher at the initial time of treatment and becomes low at a longer time (see Fig.~\ref{fig:fig8}). At t = 180 min, we observe around 70 \% MB degradation and need more time to get 100\% degradation. It means that nearly double the treatment time is required to get the same degradation of MB dye (see Fig.~\ref{fig:fig6}) when the volume of solution is doubled. It means that longer plasma treatment is required to decompose dye molecules if the volume of the solution is increased.    
\subsection{Effect of Cathode material on MB degradation} 
In the present work, a set of experiments was performed to explore the effect of cathode material on MB dye solution parameters and degradation rate. A comparative study was conducted using three different cathode materials (aluminum (Al), copper (Cu), and iron (Fe)) and common anode material. The concentration of MB solution (10 mg/L) and volume of the solution (15 ml) was kept constant in each case. We measured the parameters of the MB solution such as temperature, pH, TDS, and EC during the plasma treatment.\\
\begin{figure*} 
\subfloat{{\includegraphics[scale=0.30050]{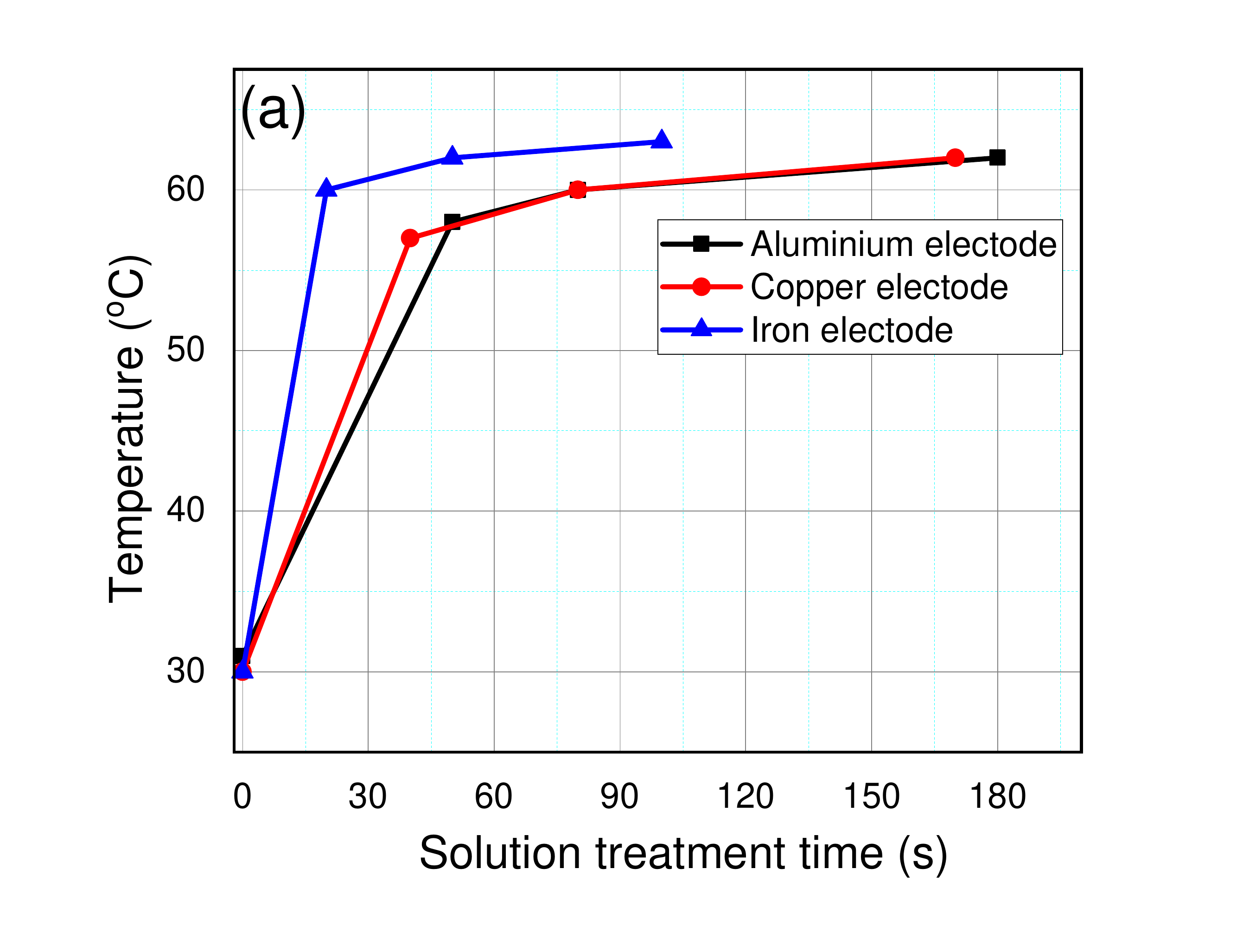}}}%
 \subfloat{{\includegraphics[scale=0.30050]{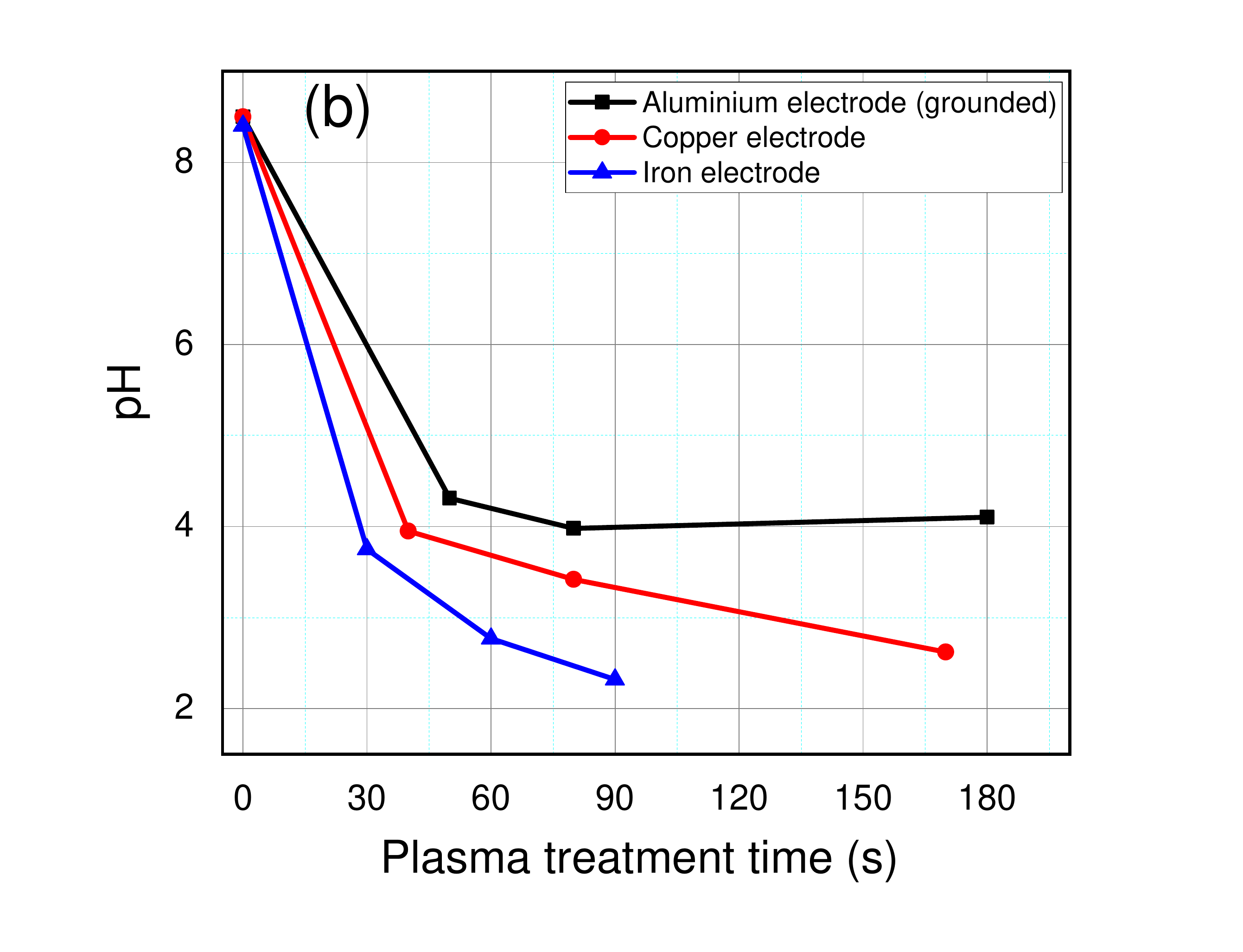}}}
 \qquad
 \subfloat{{\includegraphics[scale=0.30050]{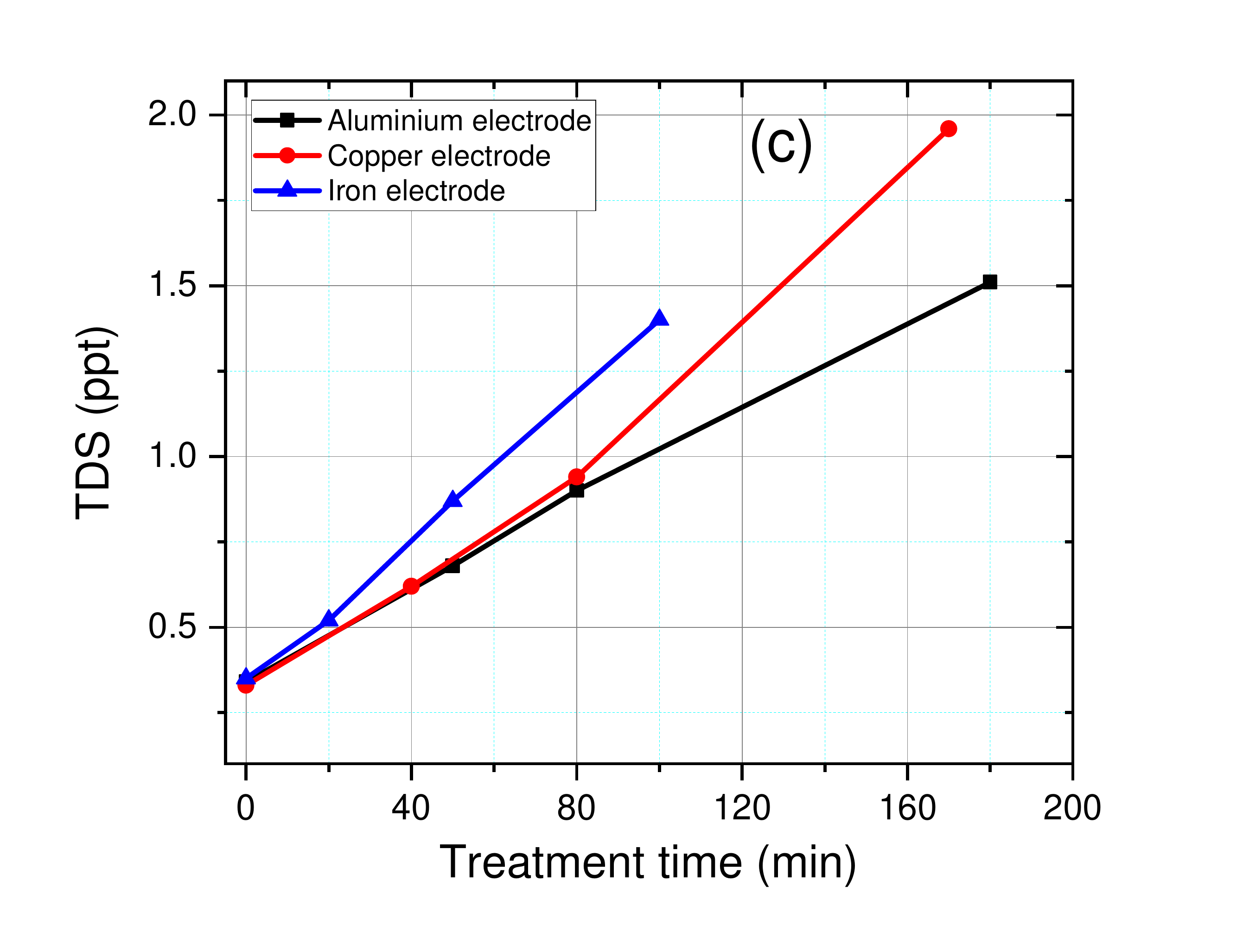}}}
 \subfloat{{\includegraphics[scale=0.30050]{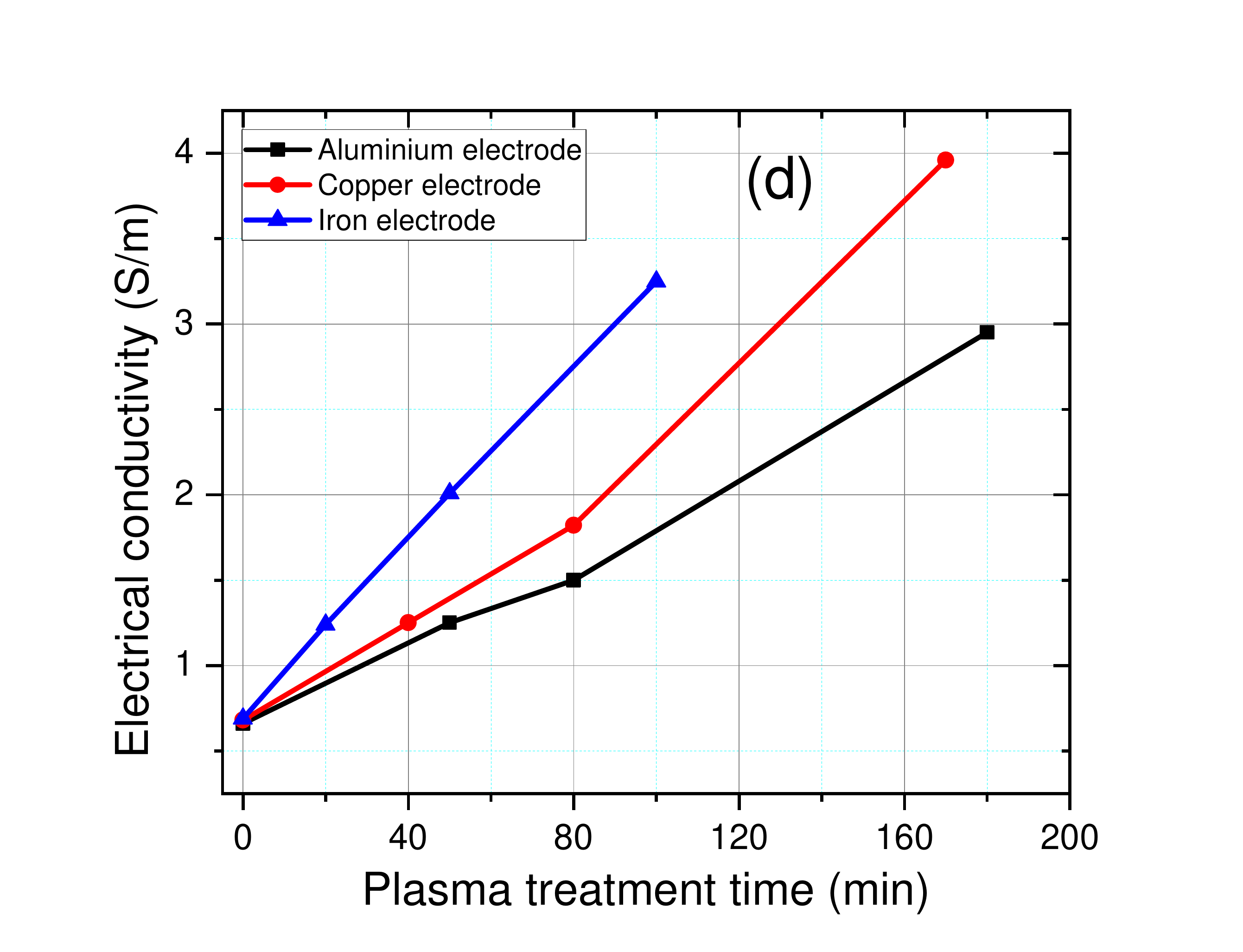}}}
\caption{\label{fig:fig9}(a)Temperature, (b) pH, (c) EC, and (d) TDS variation of MB dye solutions treated with aluminum (Al), copper (Cu) and iron (Fe) electrodes (cathode) against time. The concentration of MB solution was kept constant at 10 $mg/L$ for a comparative study. Error over the averaged value of the measured parameters is $< \pm$ 5\%} 
\end{figure*} 
\begin{figure} 
\centering
 \includegraphics[scale= 0.3200000]{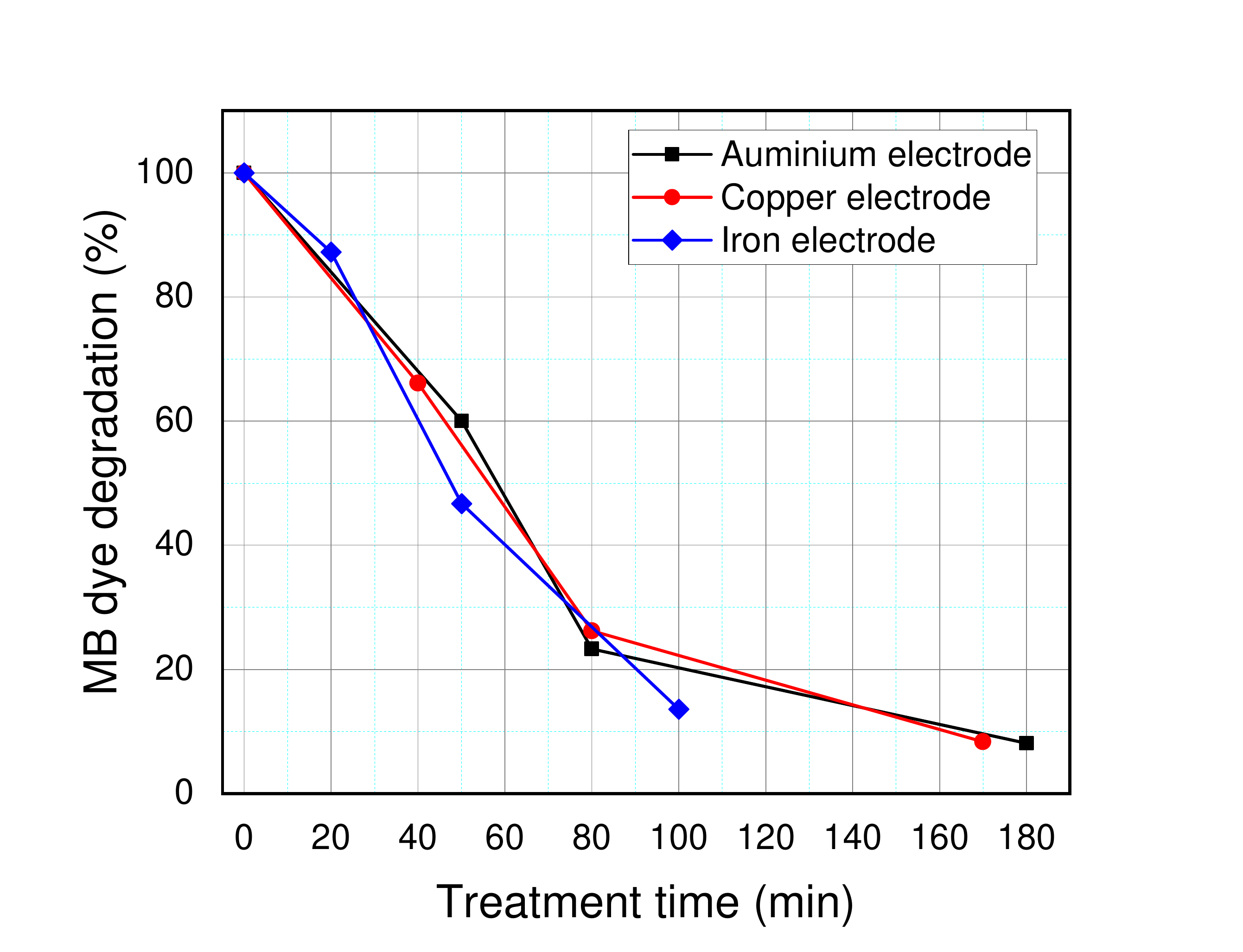}
\caption{\label{fig:fig10}Degradation (in \%) of 10 $mg/L$ concentration MB dye solution with different materials (aluminum, copper, and iron) cathode with time. Error over the plotted values (data) is $< \pm$ 5\%.}
\end{figure}
The variation of solution temperature during the plasma treatment is shown in Fig.~\ref{fig:fig9}(a). The temperature of the solution rises sharply (linearly) up to 60 minutes and then remains nearly constant for a long treatment time when aluminum or copper is used as a cathode. But in the case of the iron cathode, solution temperature rises sharply at a shorter plasma treatment time (t = 18 min) and attains a maximum temperature (see Fig.~\ref{fig:fig9}(a)) during plasma treatment. 
Fig.~\ref{fig:fig9}(b) gives the relationship between pH and plasma treatment time while three different cathodes were used. The results show that the pH decreases during the plasma treatment irrespective of cathode materials. It decreases with a higher rate during 60 min of treatment and then with a lower rate. The pH of the solution depends on the type of cathode material used in treating MB solution (see Fig.~\ref{fig:fig9})(b). The plasma-treated solution becomes more acidic in the case of an iron cathode and less in the aluminum cathode. The variation of TDS against plasma treatment time for three different cathodes is displayed in Fig.~\ref{fig:fig9}(c). The TDS increases almost linearly with plasma treatment while using any conducting material as a cathode. The highest amount of total dissolved particles was found during the use of an iron cathode at a given time. Fig.~\ref{fig:fig9}(d) shows the variation of electrical conductivity during plasma treatment. EC increases linearly (approximately) with plasma treatment time in all the cases. We observe higher electrical conductivity of the treated solution when the iron cathode is used and the solution remains less conductive in the case of the aluminum cathode at a given time. \\
For estimating the efficiency of MB degradation (in \%), plasma-treated solutions were further analyzed by a spectrometer. Before switching on the plasma source (at t = 0), the solution has 100 \% MB dye molecules but we see a reduction in MB dye molecules in all cathode material cases during the plasma treatment. The degradation rate of MB molecules is observed to be higher at the initial time of treatment and becomes lower at a longer time (see Fig.~\ref{fig:fig10}) when an aluminum or copper electrode was used as a cathode. But variation is slightly different when the iron plate was used as a cathode. MB degradation rate is found to be higher in the case of the iron cathode than the other two cathodes (Al or Cu). At about t = 100 min, we observe around 90 \% MB degradation in the iron cathode case but 75 to 80 \% when either copper or aluminum was used as cathode. There is a slight variation in MB degradation rate for copper and aluminum cathode cases as shown in Fig.~\ref{fig:fig10}. It indicates that the use of iron material as a cathode is more beneficial to degrade the MB molecules in dye-containing wastewater. A comparatively shorter plasma treatment is required to decompose the dyes in wastewater which help in making plasma technology cost-effective.  
\subsection{Effect of pH of MB solution on degradation}
\begin{figure*} 
\subfloat{{\includegraphics[scale=0.500050]{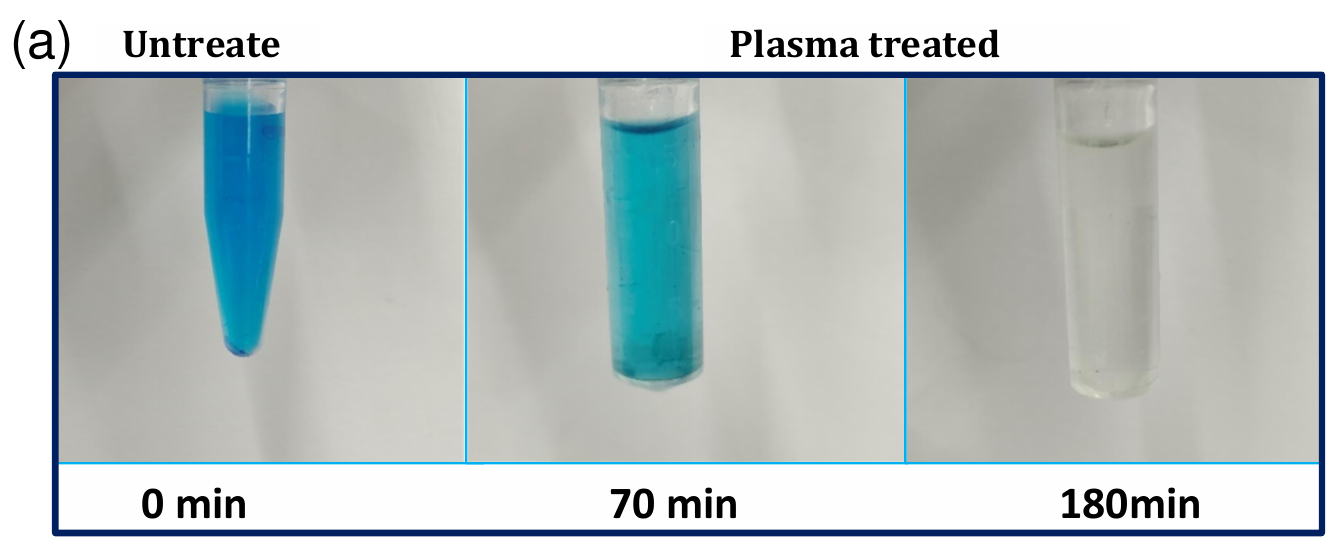}}}%
\hspace*{0.05in}
 \subfloat{{\includegraphics[scale=0.450050]{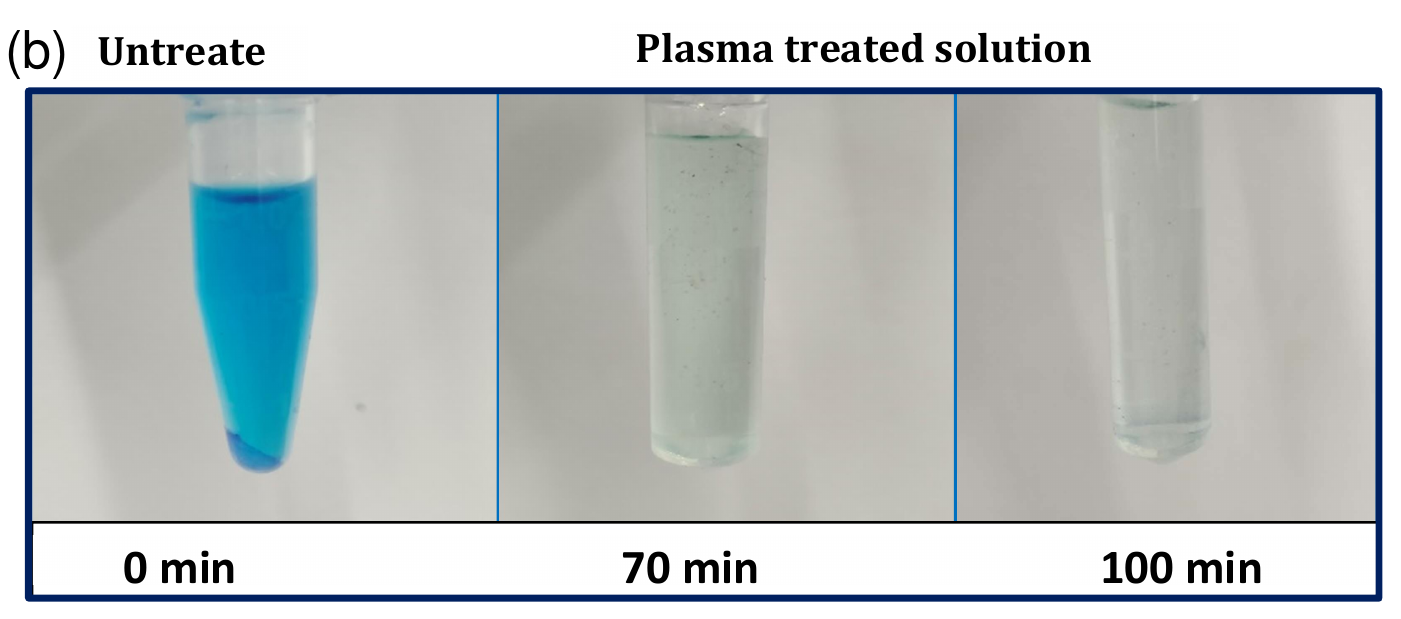}}}
 \qquad
 \subfloat{{\includegraphics[scale=0.400050]{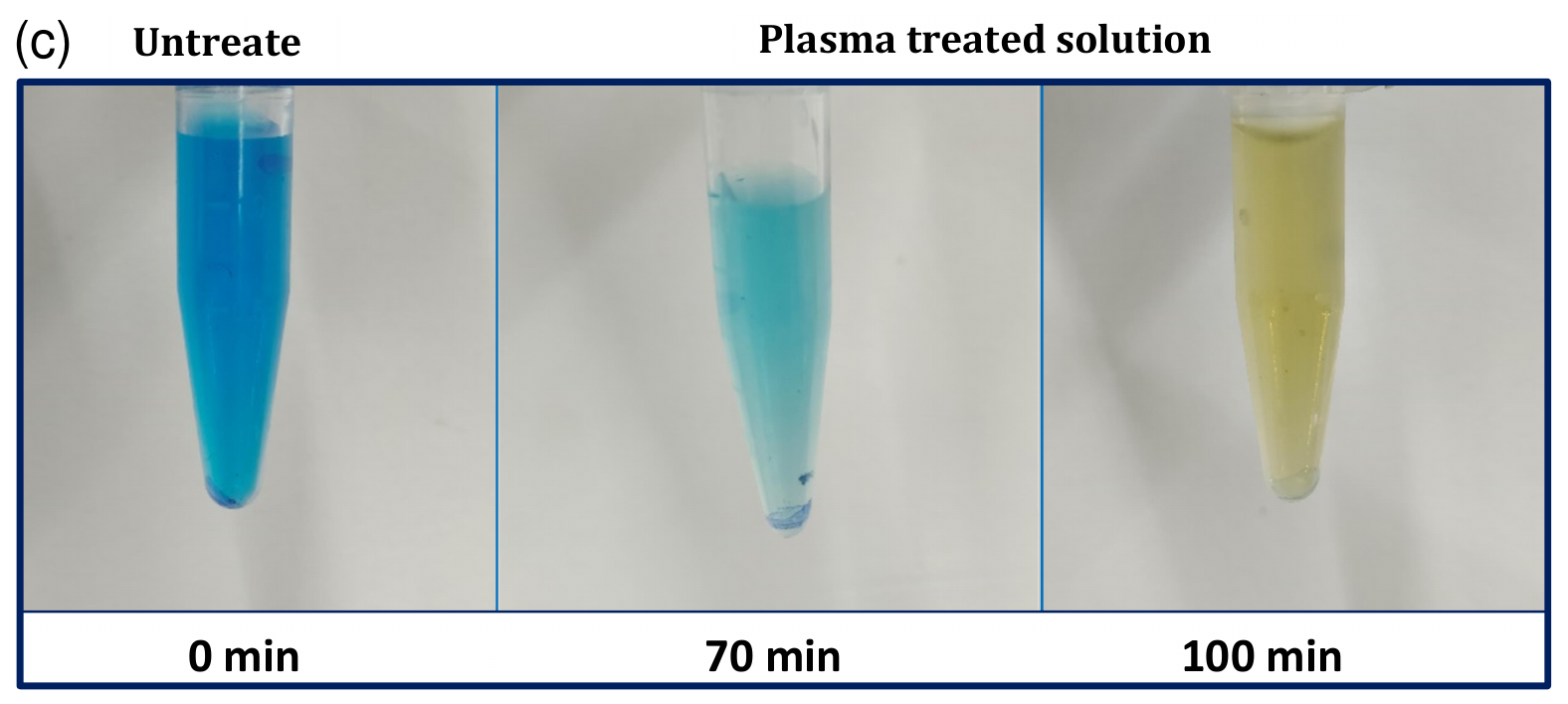}}}
 \hspace*{0.1in}
 \subfloat{{\includegraphics[scale=0.390050]{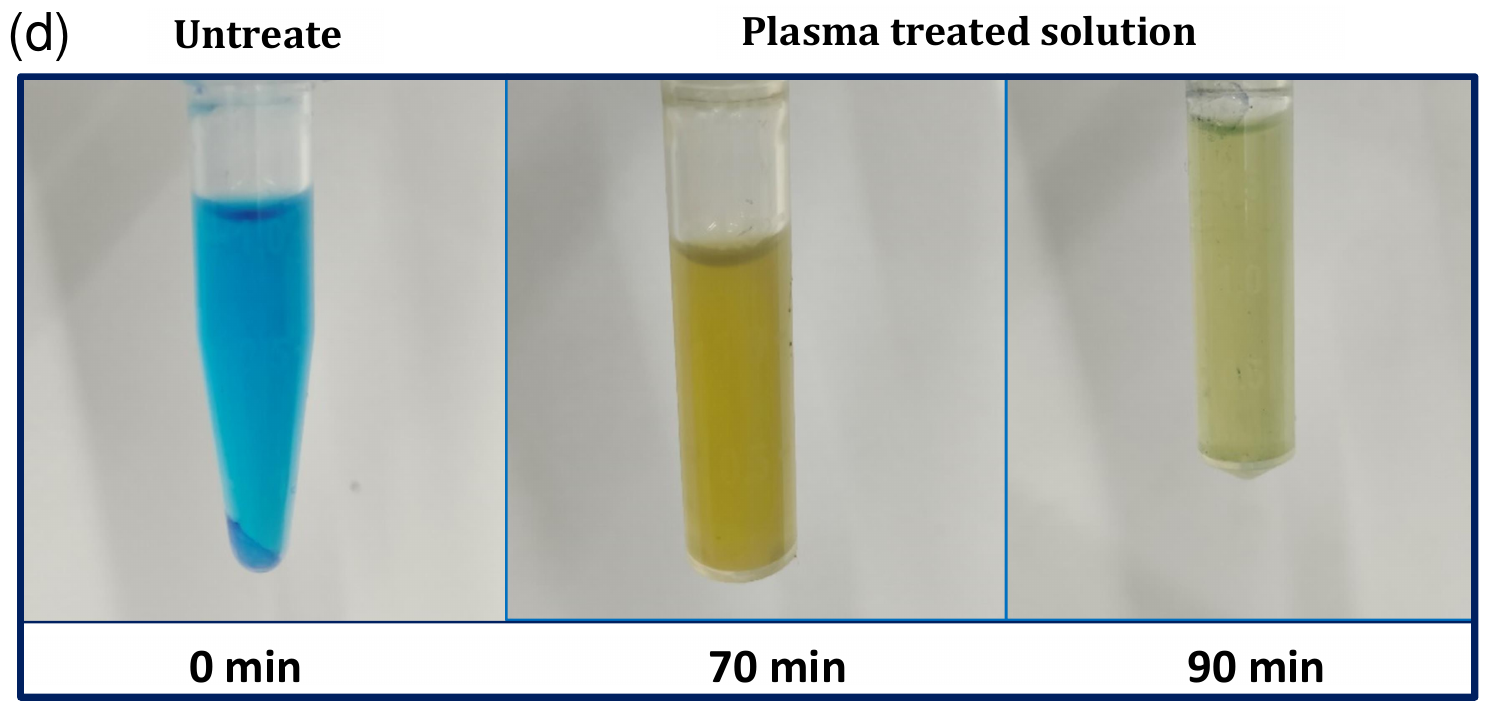}}}
 \hspace*{-0.15in}
\caption{\label{fig:fig11} Images of MB dye solution (a) Aluminium (Al) electrode and solution pH (8.5), (b)Aluminium electrode and solution pH (3.5), (c) Iron (Fe) electrode and solution pH (8.5) and Iron (Fe) electrode and solution pH (3.5) with plasma treatment time. The concentration of MB solution was 30 mg/L} 
\end{figure*}

\begin{figure} 
\centering
 \includegraphics[scale= 0.3200000]{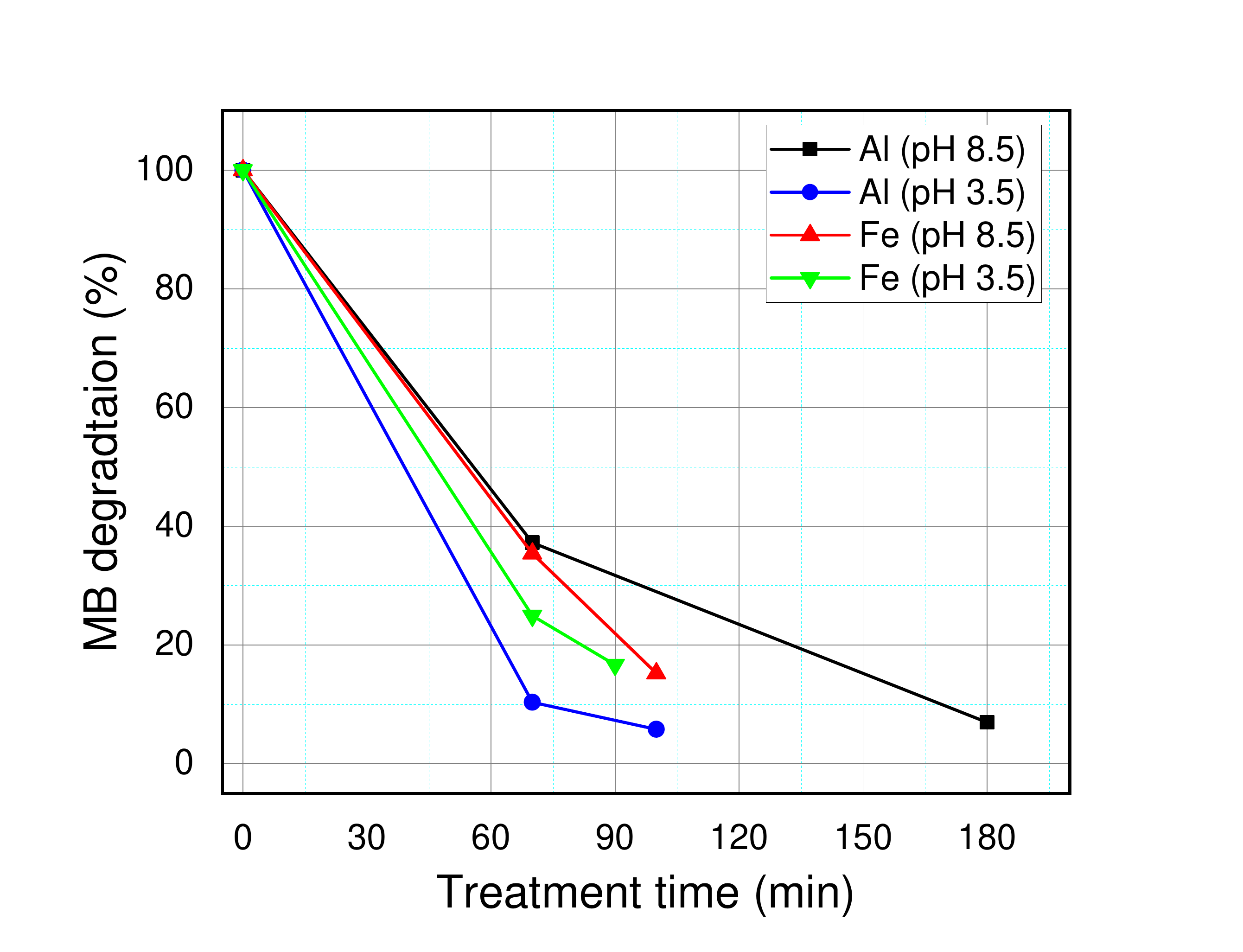}
\caption{\label{fig:fig12} Degradation (in \%) of 30 $mg/L$ concentration MB dye solution (a) Aluminium (Al) electrode and solution pH (8.5), (b)Aluminium electrode and solution pH (3.5), (c) Iron (Fe) electrode and solution pH (8.4) and Iron (Fe) electrode and solution pH (3.6) with plasma treatment time. Error over the plotted values (data) is $< \pm$ 5\%}
\end{figure}

The pH value of MB dye solution has a big influence on the oxidation processes (degradation rate) of complex organic molecules. The previous study suggests that a high pH value favors indirect oxidative reaction and in low pH (acidic) solution the direct oxidative reaction is dominant \cite{directreaction1}. Therefore, a set of experiments was performed to see the effect of the pH of the MB solution on its degradation rate during plasma treatment. The study was conducted for two different pH MB solutions with two different cathodes (aluminium and iron). We used plasma-treated MB dye solution (colorless acidic solution) to prepare the low pH MB solution. Two MB solution samples (15 mg/L) of 15 ml were treated at different times. The images of untreated and plasma-treated MB solutions of two different pH values with two different cathodes are shown in Fig.~\ref{fig:fig11}. It is noticed that low pH MB solutions in either cathode case become colorless in a short treatment time (60 to 90 min). It means that an acidic solution (pH = 3 to 3.5) takes less time in decomposing MB dye as compared to a mild basic (pH = 8 to 8.5) solution. The yellow color of the solution in the case of the iron cathode during plasma treatment is expected due to the dissolving of some iron compounds in it or the formation of some new chemical compounds. The analysis of those intermediate's chemical complexes is beyond the scope of this article. \\
The variation of pH, TDS, and EC without and with plasma treatment (aluminium cathode) is presented in Table II. Since the low pH MB solution is prepared by using a treated colorless solution, TDS and EC have higher values before treatment and increase with plasma treatment. The variation of pH is small as compared to the treated solution prepared with tap water. A similar trend of pH, TDS, and EC variation of two different pH MB solutions was noticed when experiments were performed with the iron cathode. These data are listed in Table III. \\ 
Fig.~\ref{fig:fig12} shows the degradation percentage of methylene blue with treatment time. Using an aluminum electrode with different pH MB solutions (initial) gives a large time difference for the methylene blue dye to decompose as shown in the figure. Acidic MB solution (pH = 3.5) takes less time to be colorless compared to the mild basic solution (pH = 8.5) when an aluminum cathode was used. We also observed a higher MB decomposition rate when acidic MB solution was treated with the iron cathode. Thus, it is possible to reduce the treatment time by making solutions (wastewater) acidic in nature before its plasma treatment.   

\begin{table*}
\caption{Chemical parameters of plasma treated MB dye solutions of different pH when aluminum electrode (cathode) was used.}
\begin{tabular}{|l|c|c|c|c|c|c|c|c|c|c}
\hline
\hline 
Sample &Time & pH & EC & TDS &Sample& Time & pH &EC & TDS \\
    &(min) &   & (S/m) & (ppt)& & (min) & & (S/m)& (ppt)& \\
\hline
Tap water & 0 & 8.11 & 0.61 & 0.31 &Before adding dye & 0 & 3.73 &2.23 &1.12\\   
   S1 (After adding dye) & 0 &8.19 & 0.60 & 0.30 & S1 (After adding dye) & 0& 3.71 & 2.18 & 1.09 \\
   S2 & 70 & 4.12 & 1.27 & 0.64 & S2 & 70 & 2.26 & 4.52 & 2.26 \\   
   S3 & 160 & 3.72 & 2.81 & 1.40 & S2 & 100 & 2.230 &4.84 & 2.42 \\
\hline \hline
\end{tabular}
\end{table*}
\begin{table*}
\caption{Chemical parameters of plasma treated MB dye solutions of different pH when iron electrode (cathode) was used.}
\centering 
\begin{tabular}{|l|c|c|c|c|c|c|c|c|c|c}
\hline
\hline 
Sample &Time & pH & EC & TDS &Sample& Time & pH &EC & TDS \\
    &(min) &   & (S/m) & (ppt)& & (min) & & (S/m)& (ppt)& \\
\hline
Tap water & 0 & 8.13 & 0.61 & 0.32 &Before adding dye & 0 & 3.71 &2.21 &1.15\\   
   S1 (After adding dye) & 0 &8.20 & 0.59 & 0.30 & S1 (After adding dye) & 0& 3.70 & 2.18 & 1.11 \\
   S2 & 70 & 3.32 & 1.11 & 0.56 & S2 & 70 & 2.19 & 4.12 & 2.06 \\   
   S3 & 100 & 2.21 & 3.37 & 1.69 & S2 & 90 & 2.06 &4.89 & 2.45 \\
\hline \hline
\end{tabular}
\end{table*}
\section{Conclusion} \label{sec:summary}
The experimental findings of the reported work are summarized as follows-
\begin{itemize}
    \item {} The efficiency of MB degradation depends on the concentration of the MB solution. Longer plasma treatment is required to decompose the MB molecules in a dense (high-concentration) solution. But there is not a linear relation between degradation time and MB solution concentration.
    \item{} MB degradation time is approximately proportional to the volume of treated the dye solution. Plasma treatment time will be doubled to degrade MB dye if the volume of MB solution is doubled.
    \item{} The efficiency of MB degradation strongly depends on the cathode material in atmospheric pressure discharge between the H.V. anode and the dipped cathode (grounded) in MB solution. Iron cathode reduces MB degradation time compared to aluminum or copper cathode for the same concentration and volume of the MB solution. 
    \item{} It is possible to reduce the pH of MB solution with plasma-treated water or MB solution. MB degradation time with plasma treatment decreases by reducing the pH of the MB solution before switching on the plasma source. The maximum efficiency of MB degradation can be achieved by using an aluminum cathode (electrode) in a low pH MB solution.    
\end{itemize}
It has been demonstrated in many experimental studies that air plasma at atmospheric pressure contains energetic electrons, ions, UV radiations, reactive oxygen, and nitrogen species. During the plasma--MB solution (wastewater) interaction, newly formed active oxygen species in MB solution such as hydrogen peroxide ($H_2 O_2$), hydroxyl radical ($\dot{OH}$), singlet oxygen ($1O_2$), oxygen radical ($\dot{O}$), etc along with UV radiation and energetic electrons \cite{plasmawaterinteractionreview1,plasmawaterinteractionreview2} decompose the organic contaminants (MB dye) present in the solution (wastewater). In this MB degradation process with plasma interaction, some small decomposed molecules and minerals are produced. The amount of these small decomposed molecules and other minerals decides the TDS and EC of plasma-treated MB solution. The amount of hydroxyl ions ($OH^{-}$) and hydrogen ions ($H^+$) decides the basic and acidic nature of plasma-treated MB solution respectively. The concentration of active oxidizers is expected to increase during the plasma treatment which is responsible to degrade the dissolved MB dye (or any other dye) molecules. The concentration (number) of active oxidizers in the dying solution decides the degradation efficiency of MB. Therefore, the MB degradation rate (degradation efficiency) decreases with increasing either concentration of MB dye or the volume of MB solution. In the case of an iron cathode (Fe material), MB degradation efficiency is increased compared to either aluminum or copper cathode. It is expected that some iron (Fe) ions can dissolve in the MB solution during the plasma treatment when it is used as a cathode. Therefore, we expect Fanton's reaction during the MB solution treatment with spark glow plasma. The hydroxyl radical ($\dot{OH}$) can be generated due to the reaction between hydrogen peroxide and ferrous ions. The generated hydroxyl radicals have the capability to decompose the organic compounds or dyes \cite{fantonreaction1,fantonmbdegradation1,fantonmbdegradation2}. The higher MB degradation efficiency compared to another material electrode without adding any iron compound during plasma treatment indirectly indicates the role of Fanton's reaction in the case of the iron cathode. The role of the pH of MB solution (before plasma treatment) on the efficiency of MB degradation has also been studied with copper and iron cathode. To understand the reactions to create more oxidizing agents in a low pH solution, further detailed experimental studies are required. It would be the scope of future research.    
\section{Acknowledgement} 
The authors are very grateful to Dr. Raviprakash Chandra, Dr. Alok Pandya, Dr. Gajendra Singh, Ms. Nidhi Verma, and Nilesh Patel for their assistance in chemical analysis, providing the laboratory facilities and fruitful discussion during the experiments at the Institute of Advanced Research, Gandhinagar, Gujarat, India.
\bibliography{biblography}
\end{document}